\documentclass[11pt,a4paper,reqno]{amsart}

\usepackage[latin1]{inputenc}
\usepackage[english]{babel}
\usepackage{amsmath}
\usepackage[numbers]{natbib}
\usepackage{graphicx}
\usepackage{amsmath,bm}
\usepackage{booktabs}
\usepackage{graphicx}
\usepackage{multirow}
\usepackage{setspace}
\usepackage{braket}
\usepackage[hmargin=3cm,vmargin={3cm,4cm}]{geometry}

\usepackage{etoolbox}
\newcommand*{\affaddr}[1]{#1} 
\newcommand*{\affmark}[1][*]{\textsuperscript{#1}}

\begin{document}

\title[Modelling of metals adsorption on algal-bacterial photogranules]{Multiscale modelling of heavy metals adsorption on algal-bacterial photogranules}

\author[F.~Russo et al.]{F.~Russo\protect\affmark[1] \and A.~Tenore\affmark[1] \and M.R.~Mattei\affmark[1] \and L.~Frunzo\affmark[1]}

 \maketitle

 {\footnotesize
  \begin{center}
\affaddr{\affmark[1]Department of Mathematics and Applications "Renato Caccioppoli", University of Naples Federico II, Via Cintia 1, Montesantangelo, 80126, Naples, Italy}\\
Corresponding author: M.R.~Mattei,
          \texttt{mariarosaria.mattei@unina.it}
 \end{center}}

\begin{abstract}

A multiscale mathematical model describing the genesis and ecology of algal-bacterial photogranules and the metals biosorption on their solid matrix within a sequencing batch reactor (SBR) is presented. The granular biofilm is modelled as a spherical free boundary domain with radial symmetry and a vanishing initial value. The free boundary evolution is governed by an ordinary differential equation (ODE) accounting for microbial growth, attachment and detachment phenomena. The model is based on systems of partial differential equations (PDEs) derived from mass conservation principles. Specifically, two systems of nonlinear hyperbolic PDEs model the growth of attached species and the dynamics of free adsorption sites; and two systems of quasi-linear parabolic PDEs govern the diffusive transport and conversion of nutrients and metals. The model is completed with systems of impulsive ordinary differential equations (IDEs) describing the evolution of dissolved substrates, metals, and planktonic and detached biomasses within the granular-based SBR. All main phenomena involved in the process are considered in the mathematical model. Moreover, the dual effect of metal presence on the formation process of photogranules is accounted: metal stimulates the production of EPS by sessile species and negatively affects the metabolic activities of microbial species. To describe the effects related to metal presence, a stimulation term for EPS production and an inhibition term for metal are included in all microbial kinetics. The model is used to examine the role of the microbial species and EPS in the adsorption process, and the effect of metal concentration and adsorption proprieties of biofilm components on the metal removal. Numerical results show that the model accurately describes the photogranules evolution and ecology and confirm the applicability of algal-bacterial photogranules systems for metal-rich wastewater treatment.
 
\end{abstract}

\maketitle

\section{Introduction} \label{n5.1}
 \
Increased use of metals in process industries has resulted in the production of large quantities of wastewater effluents containing high level of toxic heavy metals \cite{ahluwalia2007microbial}. Due to their non-degradable and persistent nature, tendency to accumulate, and hazardous effects on living organisms and environment, heavy metals removal represents a great challenge in the wastewater treatment field \cite{yang2015lipid, ahluwalia2007microbial, chojnacka2010biosorption, abbas2014biosorption}. The relevance of these topics in the environmental engineering has led to the development of new technologies for heavy metals removal. Methods for removing metal ions from aqueous solutions include physical, chemical, and biological processes \cite{fomina2014biosorption}. Nevertheless, conventional physical/chemical technologies, such as chemical precipitation, ion exchange, activated carbon adsorption and membrane processes, are often ineffective or expensive in the case of very low metals concentrations \cite{chojnacka2010biosorption, abbas2014biosorption, anjana2007biosorption}. A promising alternative technology for removal of heavy metals from wastewater is represented by the adsorption process and, particularly, biosorption process. Indeed, compared with conventional metals removal methods, biosorption has several advantages: utilization of renewable biomaterials; possibility to treat large volumes of wastewater; high selectivity; recovery of bound heavy metals from the biomass; no supplementation of expensive chemical reagents; low production of hazardous waste \cite{abbas2014biosorption}.

Biosorption is a complex combination of processes, consisting of the physical adherence or bonding of ions and molecules (sorbate), dissolved or suspended in a liquid phase (solvent), onto a solid surface (adsorbent) \cite{abbas2014biosorption, fomina2014biosorption,d2018free, gadd2009biosorption, papirio2017heavy}. Until now, a variety of biomaterials and microorganisms have been used as biosorbent for the removal of metals, such as algae, bacteria, fungi, and yeast \cite{abbas2014biosorption}. Such living or dead organisms are able to bind and concentrate metals, metalloids, radionuclides, and other toxic pollutants from even very dilute aqueous solutions \cite{abbas2014biosorption, fomina2014biosorption, papirio2017heavy,park2010past}. Several factors can affect the mechanism of metals biosorption: properties of the biomass (living or non-living, type of biomass, phenotype), presence of other competing ions, and environmental conditions (pH, temperature, etc.) \cite{chojnacka2010biosorption, abbas2014biosorption}. Specifically, the biomass phenotype may be considered one of the most important factors. Indeed, the use of freely-suspended microbial biosorbents has some disadvantages, including small particle size, low density, poor mechanical strength, low rigidity, difficulty in separating biomass and effluent, and poor biomass regeneration \cite{abbas2014biosorption, fomina2014biosorption, gadd2009biosorption}. For this reason, in the recent years immobilized biomass has been regarded an interesting alternative. In addition, cell agglomeration promotes the secretion of extracellular polymeric substances ($EPS$), which further contribute to microorganisms protection and metals biosorption \cite{flemming2001relevance, comte2008biosorption, liu2002characterization}.
 
The simultaneous removal of organic substances and heavy metals from wastewater is still a major engineering problem. Algal-bacterial systems are expected to have a great potential in removing organic and inorganic compounds in a single treatment step, combining high adsorption capacities of microalgae and cyanobacteria with low process costs \cite{munoz2006sequential}. Indeed, microalgae and cyanobacteria show great tendency to produce $EPS$ and high metal binding affinity. Moreover, the photosynthetic activity leads to the production of oxygen and allows the oxidation of carbon and nitrogen compounds by heterotrophic and nitrifying bacteria without external supplementation of oxygen \cite{ansari2019effects, abouhend2019growth}. In this context, self-immobilized algal-bacterial consortia in granular form, known as oxygenic photogranules (OPGs), are considered as an effective and promising technology for biosorption of inorganic pollutants and degradation of organic compounds \cite{ajiboye2021simultaneous, yang2015lipid}. In the last years, great attention has been devoted to individual removal of heavy metals \cite{yang2020enhanced, yang2021insight} or organic compounds \cite{ansari2019effects, milferstedt2017importance} in OPGs-based systems. Nevertheless, there is knowledge lack regarding their contextual removal, although these pollutants usually exist together in industrial wastewater.

In this framework, mathematical modelling represents a useful tool to explore the granulation process of OPGs and the metals adsorption on the matrix of biofilm granules. Biosorption is usually described through isotherms, which represent the equilibrium relationship between the adsorbate concentration in the liquid phase and the adsorbate concentration onto the adsorbent phase at a given temperature. For the adsorption of a single component, the most widely used isotherm model is the Langmuir-Freundlich model, which is the combination of Langmuir and Freundlich models \cite{abbas2014biosorption}. Although biosorption isotherm models have been widely recognized as efficient tools to provide a suitable description of the experimental behavior, kinetic modelling is typically preferred for practical applications and process design. Pseudo-first and pseudo-second order kinetic equations are the most widely used rate equations for the adsorption process \cite{abbas2014biosorption}. Nevertheless, more comprehensive and accurate models need to be developed to better explore the complex relationships which establish between biosorbent and sorbate. A mathematical model accounting for the biosorption process of heavy metals on the different components of a multispecies biofilm has been presented by D'Acunto et al. \cite{d2018free} in the case of planar biofilm. This mono-dimensional biofilm model is conceived in the framework of continuum mathematical modeling of biofilm growth and explicitly accounts for the diffusion and adsorption of heavy metals on the biofilm matrix. Nevertheless, none of the existing models addresses the evolution and dynamics of granular biofilms formation and the adsorption processes on their solid matrix.

In this work, we propose a mathematical model to investigate the mutual interactions between the formation of oxygenic photogranules (biosorbents) and the adsorption of heavy metals (sorbates) on their solid matrix. The \textit{de novo} granulation process of OPGs in a granular-based sequencing batch reactor (SBR) has been addressed by Tenore et al. \cite{tenore2021multiscale}. It examines all the main factors influencing the granulation process of algal-bacterial photogranules for the treatment of typical municipal wastewater. The OPGs model has been extended to explicitly account for metals diffusion from bulk liquid to biofilm and their adsorption on the matrix of biofilm granules. Following the approach proposed by Masic and Eberl \cite{mavsic2012persistence, mavsic2014modeling} in the case of one-dimensional planar biofilms, the mesoscopic granular biofilm model has been coupled to the mass balances within the macroscopic bioreactor. This multiscale approach leads to model the formation and ecology of the biofilm granules and the performances of the SBR system, considering the interaction between the granules and bulk liquid. The granular biofilm model, derived in Tenore et al. \cite{tenore2021multiscale1}, is formulated as a spherical free boundary value problem under the assumption of radial symmetry. Processes of microbial growth, attachment, and detachment are included to describe the formation and expansion of granules. The \textit{de novo} granulation process is modelled by assuming that all biomass initially present in the bioreactor is in planktonic form. Mathematically, this corresponds to consider a vanishing initial value of the granule radius, using the approach introduced by D'Acunto et al. \cite{d2019free, d2021free} in the case of planar biofilm. Attachment is modelled as a continuous flux (from the bulk liquid to the biofilm) of planktonic species, which aggregate, switch their phenotype from planktonic to sessile and initiate the granulation process. Detachment is modelled as a continuous flux (from the biofilm to the bulk liquid) proportional to the square of the granule radius. The model accounts for the first time the dynamics of the detached biomass and its influence on the biological process. Specifically, detached microbial species are modelled as a new set of variables, and are supposed to grow on soluble substrates and switch to planktonic form. Furthermore, the model includes the diffusion and consumption/production of soluble substrates, due to the metabolic activity of sessile, planktonic and detached biomasses.

The model considers the adsorption of heavy metals on the granular solid matrix. Experimental observations show that each biofilm component is characterized by the presence of specific number of adsorption sites, which are able to adsorb the contaminants present in the wastewater. For this purpose, model equations describing the variation of free binding sites, and diffusion and adsorption of metals have been here derived for the first time in the case of granular biofilm, by following the approach proposed by Tenore et al. \cite{tenore2021multiscale1}. The variation of free binding sites is assumed to depend on the biofilm growth and adsorption process, and is modelled through a system of hyperbolic partial differential equations (PDEs) \cite{d2018free}. While, the diffusion and consumption of the sorbates is described by a system of parabolic PDEs \cite{d2018free}. 

All the main components of the OPGs are accounted in the model in sessile and suspended (planktonic and detached) form: phototrophs, facultative heterotrophic bacteria, nitrifying bacteria, $EPS$ and inactive material ($EPS$ and inactive material are accounted only as sessile biomass). Since cyanobacteria (included among phototrophs) play a predominant role in the granulation of oxygenic photogranules due to their filamentous morphology \cite{ansari2019effects, abouhend2019growth, abouhend2018oxygenic}, phototrophs are assumed to have better attachment properties and to enhance the attachment of other species \cite{tenore2021multiscale}. Moreover, the model accounts the diffusion and conversion of inorganic and organic carbon, nitrate, ammonia, oxygen, and metal. The soluble substrates are involved in the metabolism of microbial species, while metal adsorbs on matrix of biofilm granules. Notably, the presence of metals in a such biological system results in a dual effect: it enhances the production of $EPS$ by sessile species \cite{naveed2019microalgal} and negatively affects the microbial metabolic activities \cite{munoz2006sequential}. This is modelled including a stimulation term for $EPS$ production and an inhibition term for metal in all microbial kinetics. Light is included in the model to consider its effects on the metabolic activity of phototrophs. Specifically, light intensity is supposed to be constant in the bulk liquid and vary within the granules due to attenuation phenomena. Various numerical studies have been performed to investigate how the metal concentration and the adsorption properties of the biofilm components may affect the evolution of the process.

The paper is organized as follows. The mathematical model is introduced and described in Section \ref{n5.2}, while the biological context is described in section \ref{n5.3}. Numerical studies and results are reported in Section \ref{n5.4} and discussed in Section \ref{n5.5}.

 \section{Mathematical Model} \label{n5.2}
\

The mathematical model simulates the biosorption process of metals within a granular-based sequencing batch reactor (SBR) with a multiscale approach. The SBR system is modelled as a batch bioreactor having a cyclic configuration, in which $N_G$ identical granules are immersed. For this purpose, two different biological compartments can be identified: the granule mesoscale and the bioreactor macroscale. The model is able to contextually describe the \textit{de novo} granulation process of granular biofilms, SBR performances, and biosorption process. The interactions between the mesoscale and macroscale are accounted in the model, by considering exchange fluxes (from/to bulk liquid and to/from biofilm) of dissolved substances (substrates, products, and metals) and biomasses (in sessile and suspended form). All main phenomena involved in the \textit{de novo} granulation process are accounted in the model: attachment process by planktonic cells; growth and decay of sessile, planktonic and detached biomasses; $EPS$ secretion; diffusion of dissolved substrates within the granule; conversion of dissolved substrates within the granules and the bulk liquid; detachment process; conversion of detached biomass into planktonic biomass. Moreover, the biosorption process of heavy metals on granule matrix is included in the model, by considering the diffusion and bioconversion of metals, and the variation of free absorption sites.

 Modelling of both the granule mesoscale and bioreactor macroscale is discussed in the following, describing the processes, assumptions, variables, equations, and initial and boundary conditions involved.

 \subsection{Granule mesoscale model} \label{n5.2.1} 
 \
The mathematical model describing the \textit{de novo} granulation process derived by Tenore et al. \cite{tenore2021multiscale1} has been here extended to model the biosorption process of heavy metals on granular biofilms matrix. The granule mesoscale consists of a fixed number of biofilm granules ($N_G$) immersed within the bulk liquid and assumed to be identical at each instant. Specifically, each granule is assumed as constituted by various particulate components (including active microbial species, extracellular polymeric substances, and inactive biomass). The granules expansion depends on growth and decay processes of the various species, attachment flux from the bulk liquid to the biofilm, and detachment flux from the biofilm to the bulk liquid. The growth of microbial species depends on the presence of nutrients necessary for their metabolic activities. The nutrients are modelled as soluble substrates able to diffuse within the granules. Granulation process is initiated by attachment of pioneering planktonic cells, while detachment phenomena lead to the loss of sessile biomass, induced by external shear forces, substrates depletion and biomass decay. Each component of the granules has a specific absorption capacity and is characterized by the presence of a certain number of free binding sites, quantified as volume fractions. The metals (sorbates) are modelled as dissolved substances, which diffuse across the granules and are subjected to absorption phenomena on the various biofilm components.  

The granular biofilm is modelled as a spherical free boundary domain under the assumption of radial symmetry. The evolution of free boundary domain is described by the variation of the granule radius $R(t)$. A vanishing initial domain ($R(0)=0$) is considered to fully model the \textit{de novo} granulation process. The center of the granule is located at $r=0$, where $r$ denotes the radial coordinate. The granule model includes $n$ microbial species in sessile form $X_i(r,t)$, $m_1$ dissolved substrates $S_j(r,t)$, $n$ free binding sites $X_{\theta_i}(r,t)$, $m_2-m_1$ heavy metals $M_j(r,t)$. All these variables are expressed in terms of concentration and modelled as functions of time $t$ and space $r$. Each microbial species is supposed to have the same biomass density $\rho$, and the same density of binding sites $\rho_{\theta}$. By dividing sessile species concentrations $X_i(r,t)$ by $\rho$ and the free binding sites concentrations $X_{\theta_i}(r,t)$ by $\rho_{\theta}$, biofilm volume fractions $f_i(r,t)$ and free binding sites volume fractions $\theta_i(r,t)$ are achieved. Notably, both $f_i(r,t)$ and $\theta_i(r,t)$ (in absence of metals adsorption) are constrained to add up to unity at each location and time ($\sum_{i=1}^n f_i$ and $\sum_{i=1}^n \theta_i$) \cite{rahman2015mixed}. 

In summary, the model components describing the granular biofilm mesoscale are:

\begin{equation} \label{5.2.1.1}
X_i, \ i=1,...,n, \ \textbf{X}=(X_1,...,X_n)
\end{equation}

\begin{equation} \label{5.2.1.2}
f_i = \frac{X_i}{\rho}, \ i=1,...,n, \ \textbf{f}=(f_1,...,f_n)
\end{equation}

\begin{equation} \label{5.2.1.3}
S_j , \ j=1, ...,m_1, \ \textbf{S}=(S_1,...,S_{m_1})
\end{equation}

\begin{equation} \label{5.2.1.4}
X_{\theta_i}, \ i=1,...,n, \ \bm{X_{\theta}}=(X_{\theta_1},...,X_{\theta_n})
\end{equation}

\begin{equation} \label{5.2.1.5}
\theta_i = \frac{X_{\theta_i}}{\rho_{\theta}}, \ i=1,...,n, \ \bm{\theta}=(\theta_1,...,\theta_n)
\end{equation}

\begin{equation} \label{5.2.1.6}
M_j , \ j=m_1+1, ...,m_2, \ \textbf{M}=(M_{m_1+1},...,M_{m_2})
\end{equation}

Based on the continuum approach introduced in Wanner and Gujer \cite{wanner1986multispecies} for one-dimensional planar biofilms, the model equations for granular biofilms were derived in Tenore et al. \cite{tenore2021multiscale1} under the assumption of radial symmetry from mass balance considerations in a differential volume of a spherical domain. 

The growth and the transport of sessile species within the granular biofilm is governed by the following system of non-linear hyperbolic partial differential equations (PDEs):

\[\frac{\partial f_i(r,t)}{\partial t} +  u(r,t)\frac{\partial f_i(r,t)}{\partial r}=r_{M,i}(r,t,{\bf f},{\bf S},{\bf M})- f_i(r,t)\sum_{i=1}^n r_{M,i}(r,t,{\bf f},{\bf S},{\bf M}),
\]
\begin{equation} \label{5.2.1.7}  
i=1,...,n, 0 \leq r \leq R(t),\ t>0,
\end{equation} 
where $r_{Mi}(r,t,{\bf f},{\bf S},{\bf M})$ is the growth rate of the $i^{th}$ sessile microbial species; and $u(r,t)$ is the biomass velocity. 

$u(r,t)$ is governed by the following equation:

\begin{equation} \label{5.2.1.8}		
\frac{\partial u(r,t)}{\partial r} = -\frac{2 u(r,t)}{r} +\sum_{i=1}^n r_{M,i}(r,t,{\bf f},{\bf S},{\bf M}), \ 0 < r \leq R(t),\ t>0.
\end{equation}

The evolution of the free boundary domain is described by the variation of the biofilm granule radius $R(t)$, according to the following equation derived from global mass balances considerations on the granule volume: 

\begin{equation} \label{5.2.1.9}
\dot R(t)  = u(R(t),t) + \sigma_a(t) - \sigma_d(t).  
\end{equation}

Attachment flux is modelled as a continuous mass flux from the bulk liquid to the granule, given by the sum of the attachment fluxes $\sigma_{a,i}(t)$ of the planktonic microbial species present in the liquid phase. The term $\sigma_{a,i}(t)$ is modelled as a linear function of the concentration of the $i^{th}$ planktonic species in the bulk liquid  \cite{d2019free, d2021free}: 

\begin{equation} \label{5.2.1.10}
\sigma_a(t)  = \sum_{i=1}^n \sigma_{a,i}(t)=\sum_{i=1}^n \frac{v_{a,i}\psi^*_i(t)}{\rho},    
\end{equation}
where $v_{a,i}$ is the attachment velocity of the $i^{th}$ planktonic species; and $\psi^*_i(t)$ is the concentration of the $i^{th}$ planktonic species within the bulk liquid.

While, detachment flux is modelled as a quadratic function of the granule radius, by following the modelling approach proposed by Abbas et al. \cite{abbas2012longtime} for the planar case. The term $\sigma_{d,i}(t)$ is modelled as the product between the detachment flux and biofilm volume fraction of the $i^{th}$ sessile biomass at the interface biofilm - bulk liquid: 

\begin{equation} \label{5.2.1.12}
\sigma_d(t)  = \sum_{i=1}^n \sigma_{d,i}(t)=\sum_{i=1}^n \lambda R^2(t) f_i(R(t),t)=\lambda R^2(t),    
\end{equation}
where $\lambda$ is the detachment coefficient and is supposed to be equal for all microbial species.

 Attachment phenomena prevail on detachment phenomena in the initial stage of the \textit{de novo} granulation process, while detachment phenomena become predominant as the granule dimension increases.
 
The diffusion and conversion of soluble substrates are governed by the following system of parabolic PDEs:

\[
\frac{\partial S_j(r,t)}{\partial t}-D_{S,j}\frac{\partial^2 S_j(r,t)}{\partial r^2} - \frac{2 D_{S,j}}{r} \frac{\partial S_j(r,t)}{\partial r}=
r_{S,j}(r,t,{\bf f},{\bf S},{\bf M}),
\]
\begin{equation} \label{5.2.1.14}
j=1,...,m_1, \ 0 < r < R(t),\ t>0,
\end{equation}
where $r_{S,j}(r,t,{\bf f},{\bf S},{\bf M})$ represents the conversion rate of the $j^{th}$ substrate; and $D_{S,j}$ denotes the diffusion coefficient in biofilm for the $j^{th}$ dissolved substrate.

Further systems of PDEs have been derived to model the variation of free binding sites and diffusion and adsorption of metals. As in the case of sessile species, the transport of free binding sites is modelled as an advective process \cite{d2018free}. Thus, the model equations governing the dynamics of the free binding sites take the following form:
 
\[	 
\frac{\partial X_{\theta_i}(r,t)}{\partial t} +\frac{1}{r^2} \frac{\partial}{\partial r}(r^2 u(r,t) X_{\theta_i}(r,t)) =  \rho_{\theta} r_{\theta,i}(r,t,{\bm X_{\theta}},{\bf M}) + \frac{\rho_{\theta}}{\rho} \mu_{M,i}(r,t,{\bf X},{\bf S},{\bf M}),
\]
\begin{equation}                                        \label{5.2.1.15}
i=1,...,n, \ 0 \leq r \leq R(t),\ t>0,
\end{equation}
 where $\mu_{M,i}(r,t,{\bf X},{\bf S},{\bf M})$ is the $i^{th}$ specific growth rate; and $r_{\theta,i}(r,t,{\bm X_{\theta}},{\bf M})$ is the consumption rate of the $i^{th}$ sessile species absorption sites. The term $\frac{\rho_{\theta}}{\rho} \mu_{M,i}(r,t,{\bf X},{\bf S},{\bf M})$ accounts for the increment of free binding sites due to the sessile biomass growth; while, their consumption is related to biosorption and decay processes, which are taken into account through $r_{\theta,i}(r,t,{\bm X_{\theta}},{\bf M})$.
 
Dividing Eq.\eqref{5.2.1.15} by $\rho_{\theta}$ and considering Eq. \eqref{5.2.1.2} and Eq. \eqref{5.2.1.5} yields:

\[
\frac{\partial \theta_i(r,t)}{\partial t} +\frac{1}{r^2} \frac{\partial}{\partial r}(r^2 u(r,t) \theta_i(r,t)) =  r_{\theta,i}(r,t,{\bm \theta},{\bf M}) + \mu_{M,i}(r,t,{\bf f},{\bf S},{\bf M}),
\]	 
\begin{equation}                                        \label{5.2.1.16}
i=1,...,n, \ 0 \leq r \leq R(t),\ t>0,
\end{equation}	

\[	 
\frac{\partial \theta_i(r,t)}{\partial t} + \theta_i(r,t) \frac{\partial u(r,t)}{\partial r} + \frac{2 u(r,t) \theta_i(r,t)}{r} + u(r,t) \frac{\partial \theta_i(r,t)}{\partial r} =
\]
\[
 = r_{\theta,i}(r,t,{\bm \theta},{\bf M}) + \mu_{M,i}(r,t,{\bf f},{\bf S},{\bf M}),
\]
\begin{equation}                                        \label{5.2.1.17}
\ i=1,...,n, \ 0 \leq r \leq R(t),\ t>0.
\end{equation} 

Substituting Eq.\eqref{5.2.1.8} into Eq.\eqref{5.2.1.17} yields:
\[
\frac{\partial \theta_i(r,t)}{\partial t} +  u(r,t)\frac{\partial \theta_i(r,t)}{\partial r}=
\]
\[	 
 r_{\theta,i}(r,t,{\bm \theta},{\bf M}) +\mu_{M,i}(r,t,{\bf f},{\bf S},{\bf M}) - \theta_i(r,t)\sum_{i=1}^nr_{M,i}(r,t,{\bf f},{\bf S},{\bf M}),
\] 
\begin{equation} \label{5.2.1.18}
 i=1,...,n, \ 0 \leq r \leq R(t),\ t>0,
\end{equation}

 As for soluble substrates, the transport of dissolved heavy metals is modelled as a diffusive process \cite{d2018free}, and it is expressed as follows:
 
 \[
\frac{\partial M_j(r,t)}{\partial t}-D_{M,j}\frac{\partial^2 M_j(r,t)}{\partial r^2} - \frac{2 D_{M,j}}{r} \frac{\partial M_j(r,t)}{\partial r}=
r_{A,j}(r,t,{\bm \theta},{\bf M}),
\]
\begin{equation} \label{5.2.1.19}
j=m_1+1,...,m_2, \ 0 < r < R(t),\ t>0,
\end{equation}
where $r_{A,j}(r,t,{\bm \theta},{\bf M})$ and $D_{M,j}$ denote the adsorption rate and diffusion coefficient of the $j^{th}$ dissolved metal within the biofilm.

 \subsection{Bioreactor macroscale model} \label{n5.2.2} 
 \
The reactor macroscale is modelled as a sequencing batch reactor in which $N_G$ granules having the same properties are immersed. Specifically, the reactor is characterized by the presence of a number of soluble substrates involved in the biological process and heavy metals taking part in the biosorption process. Besides the sessile biomass (granules), also planktonic and detached biomasses are considered in the bulk liquid. Planktonic species contribute to the genesis of the granules, while detached biomass is formed as a result of the detachment process. The modelling choice to include planktonic and detached biomass as two different variables derives from the experimental experience that the newly detached biomass has different properties from both sessile and planktonic biomass \cite{rollet2009biofilm, berlanga2014biofilm, rumbaugh2020biofilm}. Both planktonic and detached biomasses (suspended biomasses) contribute to the conversion of soluble substrates in the bulk liquid. Reconversion of detached biomass into planktonic biomass is also modelled. The bioreactor model is formulated for $n$ microbial species in planktonic form $\psi^*_i(t)$, $n$ microbial species deriving from the detachment process $\psi^*_{d_i}(t)$, $m_1$ dissolved substrates $S^*_j(t)$, and $m_2-m_1$ heavy metals $M^*_j(t)$. All these variables are expressed in terms of concentration and modelled as functions of time and not of space, since the reactor is modeled as a completely mixed reactor. An SBR is based on a sequence of treatment cycles constituted by four phases:

\begin{itemize}
\item filling phase, in which the reactor is fed with a fixed volume of wastewater;
\item reaction phase, in which  the wastewater volume is biologically treated through the biomass present in the system;
\item settling phase, which consists in the solid-liquid separation; 
\item emptying phase, in which the clarified supernatant is partially removed from the reactor. 
\end{itemize}
The filling, settling and emptying phases are supposed to be instantaneous, and the duration of the reaction phase is supposed to be the same as the cycle duration. $100$\% settling efficiency is assumed for biofilm granules, while the suspended biomass has a partial settling efficiency. Moreover, since the volume occupied by the biomass in granular and suspended form is neglected, the reactor volume is assumed to be the same as the liquid volume. The cyclic configuration of the SBR is modelled with a system of first order impulsive ordinary differential equations (IDEs) \cite{tenore2021multiscale,ferrentino2018process}. An IDE is described by three components: the continuous-time differential equation, which governs the state of the system between impulses; the impulse equation, which describes an impulsive jump and is defined by a jump function at the instant the impulse occurs; and the jump criterion, which defines a set of jump events in which the impulse equation is active.
 
In summary, the model components describing the bulk liquid are:   
\begin{equation} \label{5.2.2.1}
\psi^*_i, \ i=1,...,n, \ \bm{\psi}^*=(\psi^*_1,...,\psi^*_n),
\end{equation}

\begin{equation} \label{5.2.2.2}
\psi^*_{d_i}, \ i=1,...,n, \ \bm{\psi_{d}}^*=(\psi^*_{d_1},...,\psi^*_{d_n}),
\end{equation}

\begin{equation} \label{5.2.2.3}
S^*_j , \ j=1, ...,m_1, \ \textbf{S}^*=(S^*_1,...,S^*_{m_1}),
\end{equation}

\begin{equation} \label{5.2.2.4}
M^*_j , \ j=m_1+1, ...,m_2, \ \textbf{M}^*=(M^*_{m_1+1},...,M^*_{m_2}),
\end{equation}  
while, the system of IDEs is the following:

\[
V \dot \psi^*_i(t)= -\sigma_{a,i}(t) \rho A(t)N_G + V r^*_{\psi,i}(t,{\bm \psi^*},{\bf S^*},{\bf M^*}) +V r^*_{C,i}(t,{\bm \psi^*_{d}}),	 
\]
\begin{equation} \label{5.2.2.5}
t \in [0,T], \ t \neq t_k,i=1,...,n\,\ t>0, 
\end{equation} 

\[
V \dot \psi^*_{d_i}(t)= \sigma_{d,i}(t)\rho A(t)N_G + V r^*_{\psi_{d},i}(t,{\bm \psi^*_{d}},{\bf S^*},{\bf M^*}) - V r^*_{C,i}(t,{\bm \psi^*_{d}}),	 
\]
\begin{equation} \label{5.2.2.6}
t \in [0,T], \ t \neq t_k,i=1,...,n\,\ t>0, 
\end{equation} 

\[
V \dot S^*_j(t)=- A(t) N_G D_{S,j} \frac{\partial S_j(R(t),t)}{\partial r} +V r^*_{S,j}(t,{\bm \psi^*},{\bm \psi^*_{d}},{\bf S^*},{\bf M^*}),
\] 
\begin{equation} \label{5.2.2.7}
t \in [0,T], \ t \neq t_k, \ j=1,...,m_1,\,\ t>0,
\end{equation}

\[
V \dot M^*_j(t)=- A(t) N_G D_{M,j} \frac{\partial M_j(R(t),t)}{\partial r}, 	
\]
\begin{equation} \label{5.2.2.8}
t \in [0,T], \ t \neq t_k, j=m_1+1,...,m_2,\,\ t>0,
\end{equation}
where V is the volume of the bulk liquid; $A(t)$ is the area of the spherical granule and is equal to $4 \pi R^2(t)$; $r^*_{\psi,i}(t,{\bm \psi^*},{\bf S^*},{\bf M^*})$ and $r^*_{\psi_{d},i}(t,{\bm \psi^*_{d}},{\bf S^*},{\bf M^*})$ are the growth rates for the $i^{th}$ planktonic and detached biomass, respectively; $r^*_{S,j}(t,{\bm \psi^*},{\bm \psi^*_{d}},{\bf S^*},{\bf M^*})$ is the conversion rate for the $j^{th}$ soluble substrates; and $r^*_{C,i}(t,{\bm \psi^*_{d}})$ is the reconversion rate of the $i^{th}$ detached biomasses into planktonic form. 

The jump functions associated to Eqs. \eqref{5.2.2.5}-\eqref{5.2.2.8} are:

\begin{equation} \label{5.2.2.9}
\Delta \psi^*_i(t_k)=\psi^*_i(t^+_k)-\psi^*_i(t^-_k)= - \gamma \psi^*_i(t^-_k), \ k=1,...,h, \ i=1,...,n,
\end{equation} 
 
\begin{equation} \label{5.2.2.10}
\Delta \psi^*_{d_i}(t_k)=\psi^*_{d_i}(t^+_k)-\psi^*_{d_i}(t^-_k)= - \gamma \psi^*_{d_i}(t^-_k), \ k=1,...,h, \ i=1,...,n,
\end{equation} 
 
\begin{equation} \label{5.2.2.11}
\Delta S^*_j(t_k)=S^*_j(t^+_k)-S^*_j(t^-_k)= - \omega S^*_j(t^-_k) + \omega S^{in}_j, \ k=1,...,h, \ j=1,...,m_1,
\end{equation}

\begin{equation} \label{5.2.2.12}
\Delta M^*_j(t_k)=M^*_j(t^+_k)-M^*_j(t^-_k)= - \omega M^*_j(t^-_k) + \omega M^{in}_j, \ k=1,...,h, \ j=m_1+1,...,m_2,
\end{equation}
where $\gamma$ is the fraction of suspended biomass removed during the emptying phase; $\omega$ is the emptying/refilling ratio; $S^{in}_j$ and $M^{in}_j$ are the concentrations of the $j^{th}$ substrate and $j^{th}$ metal in the influent; $0 = t_0 <  t_1  < ... < t_h < t_{h+1} = T$, $t_{k+1} - t_k = \tau$; $\tau$ is the duration of the cycle; $\psi^*_i(t^+_k)$, $\psi^*_{d_i}(t^+_k)$, $S^*_j(t^+_k)$, $M^*_j(t^+_k)$, $\psi^*_i(t^-_k)$ ,$\psi^*_{d_i}(t^-_k)$, $S^*_j(t^-_k)$, and $M^*_j(t^-_k)$ are the right and left limits of $\psi^*_i$, $\psi^*_{d_i}$, $S^*_j$ and $M^*_j$ at time $t_k$.

Such systems of IDEs are derived from mass balance considerations and describe the dynamics of planktonic and detached biomasses, soluble substrates, and heavy metals within the bulk liquid. Equation \eqref{5.2.2.5} represents the mass balance of the $i^{th}$ microbial species in planktonic form. In particular, the mass variation over time within the bioreactor (first member) is due to the exchange flux related to the attachment process (first term of the second member), the metabolic activity in the bulk liquid (second term of the second member), and the conversion of the detached biomass into planktonic form (third term of the second member). Similarly, Eq. \eqref{5.2.2.6} represents the mass balance of the $i^{th}$ detached microbial species. In particular, the mass variation over time within the bioreactor (first member) is due to the exchange flux related to the detachment process (first term of the second member), the metabolic activity in the bulk liquid (second term of the second member), and the conversion of the detached biomass into planktonic form (third term of the second member). Obviously, the attachment flux represents a negative contribution for the planktonic biomasses, while the detachment process is a positive contribution for the detached biomasses. The conversion rate from detached to planktonic state causes two opposite contributions: positive in the equation of planktonic species and negative in the equation of detached biomasses. Eq. \eqref{5.2.2.7} represents the mass balance of the $j^{th}$ soluble substrate. In this case, the mass variation over time within the bioreactor (first member) is due to the exchange flux between the bulk liquid and the granular biofilms related to the diffusion phenomenon (first term of the second member) and its consumption and/or production occurring in the bulk liquid and mediated by the planktonic and detached biomasses (second term of second member). Lastly, Eq. \eqref{5.2.2.8} represents the mass balance of the $j^{th}$ dissolved metal. In this case, the mass variation over time within the bioreactor (first member) is due to only the exchange flux between the bulk liquid and the granular biofilms related to the diffusion phenomenon. Indeed, its consumption in the bulk liquid mediated by the planktonic and detached biomasses is neglected.
 
 \subsection{Initial and boundary conditions} \label{n5.2.3} 
 \
To integrate Eqs. \eqref{5.2.1.7}-\eqref{5.2.1.9}, \eqref{5.2.1.14}, \eqref{5.2.1.18}, \eqref{5.2.1.19}, \eqref{5.2.2.5}-\eqref{5.2.2.8}, it is necessary to specify initial and boundary conditions.
 The \textit{de novo} granulation process is modelled by coupling a vanishing initial condition to Eq. \eqref{5.2.1.9}:
 
\begin{equation} \label{5.2.3.1}
 \ R(0)= 0.
\end{equation}

The boundary condition for Eq. \eqref{5.2.1.8} is given by:

\begin{equation} \label{5.2.3.2}
   u(0,t)=0,\ t>0.
\end{equation}

The granule radius $R(t)$ represents the free boundary of the mathematical problem. Its variation, governed by Eq. \eqref{5.2.1.9}, depends on attachment $\sigma_a$ and detachment $\sigma_d$ velocity. In the initial phase, the granule radius is small and, consequently, attachment prevails on detachment. Therefore, it is $\sigma_a-\sigma_d > 0$ and the free boundary is a space-like line. During maturation, the granule dimension increases, and the detachment is the prevailing process. Thus, it is $\sigma_a-\sigma_d < 0$, and the free boundary is a time-like line. When the free boundary is a space-like line, there is a mass flux from bulk liquid to granule, and the biofilm volume fractions at the granule-bulk liquid interface are dependent on characteristics of the bulk liquid. In particular, the volume fractions of sessile biomass depend on the concentration of planktonic biomass in the bulk liquid: 

\begin{equation} \label{5.2.3.3}  
f_i(R(t),t) = \frac{v_{a,i}\psi^*_i(t)}{\sum_{i=1}^{n}v_{a,i}\psi^*_i(t)}, \ i=1,...,n,\ t>0, \ \sigma_{a}(t)-\sigma_{d}(t)>0,
\end{equation}
while, the volume fractions of the free binding sites are fixed equal to the biofilm volume fractions at the granule-bulk liquid interface:
 
\begin{equation} \label{5.2.3.4}  
\theta_i(R(t),t) = f_i(R(t),t), \ i=1,...,n,\ t>0, \ \sigma_{a}(t)-\sigma_{d}(t)>0.
\end{equation} 

When the free boundary is a time-like line, there is a mass flux from the granule to the bulk liquid. Thus, the volume fractions at the interface are regulated exclusively by the internal points of the biofilm domain and conditions \eqref{5.2.3.3} and \eqref{5.2.3.4} are not required.

For what concerns substrates and metals diffusion (Eq. \eqref{5.2.1.14} and Eq. \eqref{5.2.1.19}), a no flux condition is fixed at the granule center ($r=0$), and a Dirichlet condition is considered at the granule-bulk liquid interface ($r=R(t)$):

\begin{equation}                                        \label{5.2.3.5}
\frac{\partial S_j}{\partial r}(0,t)=0,\ S_j(R(t),t))=S^*_j(t),\ j=1,...,m_1,\ t>0,
\end{equation}

\begin{equation}                                        \label{5.2.3.6}
\frac{\partial M_j}{\partial r}(0,t)=0,\ M_j(R(t),t))=M^*_j(t),\ j=m_1+1,...,m_2,\ t>0,
\end{equation}

Note that $S^*_j(t)$ and $M^*_j(t)$ are the solutions of Eq. \eqref{5.2.2.7} and Eq. \eqref{5.2.2.8}, respectively.

Eqs. \eqref{5.2.1.7}, \eqref{5.2.1.14}, \eqref{5.2.1.18}, and \eqref{5.2.1.19} refer to the biofilm domain and do not require initial conditions, since the extension of the biofilm domain is zero at $t = 0$.
 
 Lastly, the following initial conditions are considered for Eqs. \eqref{5.2.2.5}-\eqref{5.2.2.8}:
 
\begin{equation}                                        \label{5.2.28}
\psi^*_i(0)=\psi^*_{i,0}, \ i=1,...,n,
\end{equation}

\begin{equation}                                        \label{5.2.29}
\psi^*_{d_i}(0)=\psi^*_{d_{i,0}}, \ i=1,...,n,
\end{equation}
 	
\begin{equation}                                        \label{5.2.30}
S^*_j(0)=S^*_{j,0},\ j=1,...,m_1,
\end{equation}
 
\begin{equation}                                        \label{5.2.18}
M^*_j(0)=M^*_{j,0},\ j=m_1+1,...,m_2,
\end{equation}
where $\psi^*_{i,0}$, $\psi^*_{d_i,0}$, $S^*_{j,0}$, and $M^*_{j,0}$ are the initial concentrations of the $i^{th}$ planktonic and detached biomass, and the $j^{th}$ soluble substrate and dissolved metal within the bulk liquid, respectively. 

\section{Biochemical framework: OPGs granulation and adsorption processes} \label{n5.3}
 \
The mathematical model described above simulates the biosorption process of metals on the matrix of biofilm granules, occurring in a granular-based SBR system, and is able to contextually describe granules genesis and ecology, bioreactor performances, and adsorption process of inorganic compounds. In this work, the model is applied to study the ecology of OPGs and the adsorption process of a generic metal on their solid matrix. 
 
\subsection{Metabolic kinetics of OPGs} \label{n5.3.1}
\
All main biological processes involved in the OPGs lifecycle are included in the mathematical model. For this purpose, phototrophs $PH$, heterotrophic bacteria $H$, and nitrifying bacteria $N$ are taken into account as active microbial species. While, the following soluble substrates are considered: inorganic carbon $IC$, organic carbon $DOC$, nitrate $NO_3$, ammonia $NH_3$, and dissolved oxygen $O_2$.

The growth metabolism of phototrophs is affected by light. Two different processes of phototrophic growth in presence of light are taken into account, based on the available nitrogen source. In presence of $NH_3$, phototrophs carry out photosynthesis, consuming $IC$ and $NH_3$ and producing $O_2$ and $DOC$. In absence or shortage of ammonia, phototrophs can grow by using $NO_3$ as nitrogen source. Furthermore, the model takes into account the inhibition induced by the presence of $O_2$ on the photosynthetic activity. In absence of light, $DOC$, $O_2$, and $NH_3$ are consumed by the phototrophs, which produce $IC$. Heterotrophic bacteria use $DOC$ as a source of carbon and energy, and produce inorganic carbon $IC$. They are assumed to grow under aerobic condition directly using $O_2$, as well as anoxic condition using $NO_3$ as oxygen source (denitrification process). As in the previous case, this aspect is modelled using an inhibition term for oxygen in the nitrate-based heterotrophic growth kinetic \cite{wolf2007kinetic}. Nitrifying bacteria include ammonia-oxidizing bacteria and nitrite-oxidizing bacteria. For this reason, they are responsible for $NH_3$ conversion into $NO_2$, and the subsequent $NO_2$ conversion into $NO_3$. The same biological processes are supposed to occur in the bulk liquid, where planktonic and detached biomasses consume or produce the $j^{th}$ soluble substrate. The mathematical model considers the production of $EPS$ and inactive material only in sessile form. Indeed, the $EPS$ production by suspended biomass has been neglected because it is much lower than sessile production  \cite{tenore2021multiscale}, as well as the production of suspended inactive biomass that does not play any role in the biological process. Moreover, phototrophs are regarded as the main $EPS$ contributors \cite{naveed2019microalgal}, and this aspect has been considered in the model by adopting different values of $EPS$ fraction produced by the microbial species. 

\subsection{Adsorption process} \label{n5.3.2}
\
 Compared to conventional physical/chemical technologies, biosorption is effective and less expensive when the metal concentration is less than $100 \ mg \ L^{-1}$ \cite{yang2015lipid,ahluwalia2007microbial, chojnacka2010biosorption, abbas2014biosorption, anjana2007biosorption}. Both living and dead (metabolically inactive) biological materials are able to adsorb toxic heavy metals, as various functional groups are found on their cell wall offering strong attraction forces for the metal ions and providing high metal removal efficiency \cite{abbas2014biosorption}. Specifically, extracellular substances produced by microorganisms have a crucial role in biosorption of metals \cite{liu2002characterization} and are considered the major potential agents in biosorption processes \cite{naveed2019microalgal}. In metal-stressed conditions microorganisms are induced to produce a higher amount of $EPS$, increasing the adsorption potential of the microbial consortium \cite{naveed2019microalgal, singh2006biofilms}. Moreover, microorganisms not only regulate the synthesis of $EPS$ in response to toxic elements, but also increase $EPS$ adsorption capacities \cite{naveed2019microalgal}. These aspects are included in the model by adopting a higher adsorption constant for $EPS$ and considering a stimulation term for $EPS$ production in all microbial kinetics. Also phototrophs and inactive material play an important role in adsorption processes, as they show high metals removal efficiency and can achieve more effective biosorption of metals than bacteria and fungi \cite{singh2006biofilms}. Indeed, metal accumulation capacity of phototrophs is comparable or sometimes higher than chemical sorbents \cite{mehta2005use}, and, in addition, as mentioned before they are the main $EPS$ producers \cite{naveed2019microalgal}. The use of dead biomass could be a preferred alternative, as it offers high metals adsorption capacity, easy recovery of biosorbed metals, absence of toxicity limitations and nutrients requirements for growth \cite{abbas2014biosorption,fomina2014biosorption}. However, in the case in which the solvent consists of industrial wastewater rich in metals, organic and nitrogen compounds, the problem related to the nutrients requirement is overcome and the utilization of algal-bacterial biomass allows to combine the advantages of $EPS$, microalgae and inactive material. Lastly, metals adsorption by suspended biomasses is neglected. Indeed, populations of planktonic and biofilm cells adsorb metals in different ways \cite{harrison2007multimetal}, and it is experimentally proved that immobilized bacterial cells have much higher biosorption capacities than suspended cells \cite{rani2010comparative}. Moreover, the use of freely-suspended microbial biosorbents has further disadvantages including small particle size, low density, poor mechanical strength, and little rigidity, while the use of biofilms minimize these disadvantages \cite{fomina2014biosorption}.  

\subsection{Modelling of heavy metals adsorption on OPGs} \label{n5.3.3}
\
In summary, the following variables are included in the model:
\begin{itemize}
\item Granule variables:
\begin{itemize}
\item Five sessile microbial species: phototrophs $f_{PH}(r,t)$, heterotrophic bacteria $f_{H}(r,t)$, nitrifying bacteria $f_{N}(r,t)$, $EPS$ $f_{EPS}(r,t)$, and inactive biomass $f_{I}(r,t)$. 
\item Five soluble compounds: inorganic carbon $S_{IC}(r,t)$, organic carbon $S_{DOC}(r,t)$, nitrate $S_{NO_3}(r,t)$, ammonia $S_{NH_3}(r,t)$, and dissolved oxygen $S_{O_2}(r,t)$.
\item Five fractions of free binding sites related to: phototrophs $\theta_{PH}(r,t)$, heterotrophic bacteria $\theta_{H}(r,t)$, nitrifying bacteria $\theta_{N}(r,t)$, $EPS$ $\theta_{EPS}(r,t)$, and inactive biomass $\theta_{I}(r,t)$.
\item One metal: $M(r,t)$.
\end{itemize}
\item SBR variables:
\begin{itemize}
\item Three planktonic microbial species: phototrophs $\psi^*_{PH}(t)$, heterotrophic bacteria $\psi^*_{H}(t)$, and nitrifying bacteria $\psi^*_{N}(t)$.
\item Three microbial species deriving from biofilm detachment: phototrophs $\psi^*_{d_{PH}}(t)$, heterotrophic bacteria $\psi^*_{d_H}(t)$, and nitrifying bacteria $\psi^*_{d_N}(t)$.
\item Five soluble compounds: inorganic carbon $S^*_{IC}(t)$, organic carbon $S^*_{DOC}(t)$, nitrate $S^*_{NO_3}(t)$, ammonia $S^*_{NH_3}(t)$, and dissolved oxygen $S^*_{O_2}(t)$.
\item One metal: $M^*(t)$
\end{itemize}
\end{itemize}
 
 In order to account the light dependency of the phototrophic metabolism, light intensity is included as a model variable: $I(r,t)$. I is assumed to be a piecewise-constant function in the bioreactor to simulate the day-night cycle, while it varies across the granule radius due to attenuation phenomena, according to the Lambert-Beer law:
 
\begin{equation} \label{5.3.3.1}
I(r,t) = I_0 e^{-k_{tot}(R(t)-r)\rho}, 0 \leq r \leq R(t),\ t>0,
\end{equation}
where $I_0$ is the light intensity in the bioreactor and $k_{tot}$ is the light attenuation coefficient \cite{tenore2021modelling}. Also the negative effect of photoinhibition is accounted in the model \cite{tenore2021multiscale}. Indeed, excess light can photoinhibit and slow down photosynthesis activity of phototrophs \cite{long1994photoinhibition}. In accordance with \cite{steele1962environmental}, such phenomena is modelled by using the following optimum type expression:

\begin{equation} \label{5.3.3.2}
\phi_{I}(r,t) = \frac{I(r,t)}{I_{opt}} \ e^{1-\frac{I(r,t)}{I_{opt}}}, \ 0 \leq r \leq R(t),\ t>0,
\end{equation}
where $I_{opt}$ is the optimum light intensity for phototrophs.

The OPGs granulation process is governed by phototrophs, in planktonic form which are able to aggregate and enclose non-phototrophic biomass in a rigid and spherical structure thanks to their filamentous morphology \cite{ansari2019effects, milferstedt2017importance}. As proposed by Tenore et al. \cite{tenore2021multiscale}, this aspect is taken into account assuming that the attachment velocities of heterotrophic and nitrifying bacteria are functions of the planktonic phototrophs concentration within the bioreactor: 

\begin{equation} \label{5.3.3.3}
\sigma_{a,i}(t)=\frac{v_{a,i} \psi^*_{i}(t)}{\rho}, \ i \in \{PH, H, N \},
\end{equation}
 
\begin{equation} \label{5.3.3.4}
v_{a,i}= v^0_{a,i}, \ i \in \{PH \},
\end{equation} 
 
\begin{equation} \label{5.3.3.4}
v_{a,i}(\psi^*_{PH}(t))= \frac{v^0_{a,i} \psi^*_{PH}(t)}{K_{PH}+\psi^*_{PH}(t)}, \ i \in \{H, N \},
\end{equation}
where $v^0_{a,i}$ is the maximum attachment velocity of the $i^{th}$ suspended species and $K_{PH}$ is the cyanobacteria half saturation constant on the attachment of heterotrophs and nitrifiers.

All reaction terms of the model are reported below. The sessile biomasses growth rates $r_{M,i}$ and conversion rates for soluble substrates within the biofilm $r_{S,j}$, reported in Eq. \eqref{5.2.1.7}, Eq. \eqref{5.2.1.8}, and Eq. \eqref{5.2.1.14} are modelled as follows:

 \begin{equation}                                        \label{5.3.3.5}
r_{M,i} =  \sum_k \alpha_{i,k} \ \nu_{k}, \ i \in \{PH, H, N,EPS,I \}, \ k=1,...,m,
 \end{equation}
  
 \begin{equation}                                        \label{5.3.3.6}
r_{S,j} =  \sum_k \beta_{j,k} \ \nu_{k}, \ j \in \{IC, DOC, NO_3, NH_3, O_2 \}, \ k=1,...,m,
 \end{equation}
 where $m$ denotes the number of biological processes occurring in the biofilm and accounted in the mathematical model; $\alpha_{i,k}$ is the stoichiometric coefficient of the $i^{th}$ biofilm component referred to the $k^{th}$ biological process within the biofilm (Table \ref{t5.1}); $\beta_{j,k}$ is the stoichiometric coefficient of the $j^{th}$ soluble substrate referred to the $k^{th}$ biological process within the biofilm (Table \ref{t5.2}); $\nu_{k}$ represents the kinetic rate of the $k^{th}$ biological process within the biofilm (Table \ref{t5.3}). Moreover, the conversion rates of planktonic biomasses $r^*_{\psi,j}$, detached biomasses $r^*_{\psi_{d_i}}$ and soluble substrates $r^*_{S,j}$ within the bulk liquid, reported in Eqs. \eqref{5.2.2.5}-\eqref{5.2.2.7} are defined as:
 
  \begin{equation}                                        \label{5.3.3.7}
r^*_{\psi,i} = \sum_k \alpha^*_{i,k} \ \nu^*_{k}, \ i \in \{PH, H, N \}, \ k=1,...,m,
 \end{equation}
 
   \begin{equation}                                        \label{5.3.3.8}
r^*_{\psi_{d_i}} =  \sum_k \bar{\alpha}^*_{i,k} \ \nu^*_{k}, \ i \in \{PH, H, N \}, \ k=1,...,m,
 \end{equation}

  \begin{equation}                                        \label{5.3.3.9}
r^*_{S,j} =  \sum_k \beta^*_{j,k} \ \nu^*_{k}, \ j \in \{IC, DOC, NO_3, NH_3 \}, \ k=1,...,m,
 \end{equation}
 
   \begin{equation}                                        \label{5.3.3.10}
r^*_{S,j} =  \sum_k \beta^*_{j,k} \ \nu^*_{k} + k_{La} (S_{j,sat} - S_{j}^*), \ j \in \{O_2 \}, \ k=1,...,m,
 \end{equation}
 where $m$ denotes the number of biological processes occurring in the bulk liquid and accounted in the mathematical model; $k_{La}$ and $S_{O_2,sat}$ are the oxygen mass transfer coefficient and the oxygen saturation concentration in the bulk liquid, respectively; $\alpha^*_{i,k}$ is the stoichiometric coefficient of the $i^{th}$ planktonic species referred to the $k^{th}$ biological process within the bulk liquid (Table \ref{t5.4}); $\bar{\alpha}^*_{i,k}$ is the stoichiometric coefficient of the $i^{th}$ detached species referred to the $k^{th}$ biological process within the bulk liquid (Table \ref{t5.4}); $\beta^*_{j,k}$ is the stoichiometric coefficient of the $j^{th}$ soluble substrate referred for the $k^{th}$ biological process within the bulk liquid (Table \ref{t5.5}); $\nu^*_{k}$ represents the kinetic rate of the $k^{th}$ biological process within the bulk liquid (Table \ref{t5.6}).
 
 Detached biomass has different characteristics from both sessile and planktonic biomasses \cite{rollet2009biofilm, berlanga2014biofilm, rumbaugh2020biofilm}. Experimental observations suggest that the surface properties of detached cells clearly differ from those of planktonic and sessile cells for at least the first $48 \ h$ after detachment \cite{berlanga2014biofilm}. For this reason, the reconversion rate of the $i^{th}$ detached species into planktonic form, reported in Eq. \eqref{5.2.2.5} and Eq. \eqref{5.2.2.6}, is modelled as follows:
  
 \begin{equation} \label{5.3.3.11}
r^*_{C,i} = K_C\psi_{d_i}^*(t),
 \end{equation} 
 where $K_C$ is the conversion constant from detached to planktonic form.  
 
Regarding the adsorption process, the consumption rate of the free binding sites $r_{\theta,i}$ and the specific growth rate of sessile species $\mu_{M,i}$ in Eq. \eqref{5.2.1.18}, and the adsorption rate of the dissolved metal $r_{A}$ in Eq. \eqref{5.2.1.19} are listed below:
 
 \begin{equation}                                        \label{5.3.3.12}
r_{\theta,i} = - (k_{ads,i} M+k_{d,i}) \theta_{i}, \ i \in \{PH, H, N,EPS,I \},
 \end{equation}
 
  \begin{equation}                                        \label{5.3.3.12.2}
\mu_{M,i} =  \sum_k \alpha_{i,k} \ \nu_{k}, \ i \in \{PH, H, N,EPS,I \}, \ k=1,...,\bar m,
 \end{equation}
 
  \begin{equation}                                        \label{5.3.3.13}
r_{A} = - Y_{ads,i}\rho_{\theta} k_{ads,i} M \theta_{i}, \ i \in \{PH, H, N,EPS,I \},
 \end{equation} 
 where $\bar m$ denotes the number of growth processes occurring in the biofilm and accounted in the mathematical model; $k_{d,i}$ is the decay-inactivation rate for the $i^{th}$ microbial species; $Y_{ads,i}$ and $k_{ads,i}$ represent the biosorption yield and the adsorption kinetic constant of the $i^{th}$ microbial species; $\rho_{\theta}$ is the density of the binding sites.

As mentioned above, in metal-stressed conditions microorganisms regulate the synthesis of $EPS$ and are induced to produce more. To include this aspect in the model, $EPS$ fraction produced by each the microbial species is modelled as function of the metal concentration:

\begin{equation} \label{5.3.3.14}
K_{EPS,i}= \tilde{K}_{EPS,i} \ \Big(1+ \frac{M}{K_{s,i}+M}\Big),
\end{equation}
where $\tilde{K}_{EPS,i}$ and $K_{s,i}$ are the $EPS$ fraction produced by the $i^{th}$ microbial species in absence of toxic pollutants and the stimulation constant for $EPS$ of the $i^{th}$ microbial species. The model takes into account also the toxic effect of metals on microbial metabolic processes, by considering an inhibition term in all microbial kinetics:

\begin{equation} \label{5.3.3.15}
I_M= \frac{K^{in}_{M}}{K^{in}_{M}+M},
\end{equation}
where $K^{in}_{M}$ is the inhibition coefficient for the generic heavy metal.

The values used for all stoichiometric and kinetic parameters are reported in Table \ref{t5.7}.
 
 \clearpage
\begin{table}[ht!]
\centering
\rotatebox{90}{%
\makebox[0pt][c]{\parbox{3.0\textwidth}{%
    \begin{minipage}[b]{1\vsize}
    \begin{center}  
        \begin{tabular}{|c|c|c|c|c|c|c|c|}
\hline
{\textbf{}} & {$\bf{x_{PH}}$} & {$\bf{x_{H}}$} & {$\bf{x_{N}}$} & {$\bf{x_{EPS}}$} & {$\bf{x_{I}}$} & {\textbf{Rate}} \\
\hline
\tiny{1}& \tiny{$1-k_{EPS,PH}$}  & & & \tiny{$k_{EPS,PH}$} & & \tiny{$\nu_{1}$}\\
\hline
\tiny{2} & \tiny{$1-k_{EPS,PH}$}  & & & \tiny{$k_{EPS,PH}$} & &\tiny{$\nu_{2}$} \\
\hline
\tiny{3} & \tiny{$1-k_{EPS,PH}$}  & & & \tiny{$k_{EPS,PH}$} & & \tiny{$\nu_{3}$}\\
\hline
\tiny{4} & & \tiny{$1-k_{EPS,H}$} & & \tiny{$k_{EPS,H}$} & & \tiny{$\nu_{4}$} \\
\hline
\tiny{5} & & \tiny{$1-k_{EPS,H}$} & & \tiny{$k_{EPS,H}$} & & \tiny{$\nu_{5}$} \\
\hline
\tiny{6} & & &\tiny{$1-k_{EPS,N}$} & \tiny{$k_{EPS,N}$} & & \tiny{$\nu_{6}$} \\
\hline
\tiny{7} & \tiny{-1} &  & & & \tiny{1} & \tiny{$\nu_{7}$} \\
\hline
\tiny{8} & & \tiny{-1} & & & \tiny{1} & \tiny{$\nu_{8}$} \\
\hline 
\tiny{9} & & & \tiny{-1} & & \tiny{1} & \tiny{$\nu_{9}$} \\
\hline
        \end{tabular}
        \caption{Biochemical rate coefficients ($\alpha_{i,k}$) of the biological processes within the biofilm.}
        \label{t5.1}
         \end{center}
    \end{minipage}
    \vfill
    \vspace{20mm}
        \begin{minipage}[b]{1\vsize}
        \begin{center}
        \begin{tabular}{|c|c|c|c|c|c|c|}
 \hline
{\textbf{}} & {$\bf{S_{IC}}$} & {$\bf{S_{DOC}}$} & {$\bf{S_{NO_3}}$} & {$\bf{S_{NH_3}}$} & {$\bf{S_{O_2}}$} & {\textbf{Rate}}
 \\
 \hline
 \tiny{1} & \tiny{$-\frac{k_{EPS,PH}+(1-k_{EPS,PH})(k_{DOC}+1.0025)}{32}$} & \tiny{$k_{DOC}(1-k_{EPS,PH})$} & \tiny{$-\frac{0.1704}{32}(1-k_{EPS,PH})$} & & \tiny{$\frac{k_{EPS,PH}+(1-k_{EPS,PH})(k_{DOC}+1)}{32}$} & \tiny{$\nu_{1}$} \\
 \hline
 \tiny{2} & \tiny{$-\frac{k_{EPS,PH}+(1-k_{EPS,PH})(k_{DOC}+1.0025)}{32}$} & \tiny{$k_{DOC}(1-k_{EPS,PH})$} & & \tiny{$-\frac{0.1704}{32}(1-k_{EPS,PH})$} & \tiny{$\frac{k_{EPS,PH}+(1-k_{EPS,PH})(k_{DOC}+1.3409)}{32}$} & \tiny{$\nu_{2}$} \\
 \hline
 \tiny{3} & \tiny{$\frac{1-1.0025 \ Y_{DOC}}{32 \ Y_{DOC}}$} & \tiny{$-\frac{1}{Y_{DOC}}$} & & \tiny{$-\frac{0.1704}{32 \ Y_{DOC}}$} & \tiny{$-\frac{1-Y_{DOC}}{32 \ Y_{DOC}}$} & \tiny{$\nu_{3}$}\\
 \hline
 \tiny{4} & \tiny{$\frac{1}{32 \ Y_H}-0.02976$} & \tiny{$-\frac{1}{Y_H}$} & & \tiny{$-\frac{0.2}{33.6}$} & \tiny{$-\frac{1}{32 \ Y_H}+0.03125$} & \tiny{$\nu_{4}$} \\
 \hline
 \tiny{5} & \tiny{$\frac{1}{32 \ Y_H}-0.02976$} & \tiny{$-\frac{1}{Y_H}$} & \tiny{$-\frac{0.8}{32 \ Y_H}+0.02857$}  & \tiny{$-\frac{0.2}{33.6}$} &  & \tiny{$\nu_{5}$} \\
 \hline
 \tiny{6} & \tiny{$-\frac{1}{33.6}$} & & \tiny{$\frac{1}{14 \ Y_N}$} & \tiny{$-(0.00593 + \frac{1}{14 \ Y_N})$} & \tiny{$-\frac{1}{7 \ Y_N}+\frac{1}{32}$} & \tiny{$\nu_{6}$} \\
 \hline
  \tiny{7} & & & & & & \tiny{$\nu_{7}$} \\
 \hline
  \tiny{8} & & & & & & \tiny{$\nu_{8}$} \\
 \hline
  \tiny{9} & & & & & & \tiny{$\nu_{9}$} \\
 \hline
        \end{tabular}
        \caption{Biochemical rate coefficients ($\beta_{j,k}$) of the biological processes within the biofilm.}
        \label{t5.2}
        \end{center}
    \end{minipage}%
}}%
}
\end{table}  
\clearpage

\clearpage
\begin{table}[ht!]
\begin{footnotesize}
\centering
\rotatebox{90}{%
\makebox[0pt][c]{\parbox{3.0\textwidth}{%
    \begin{minipage}[b]{1\vsize}
    \begin{center}  
    \vspace{25mm}
        \begin{tabular}{|l|l|l|}
\hline
{\textbf{}} & {\textbf{Process}} & {\textbf{Kinetic rate expression}} 
 \\
 \hline
 \tiny{1} & Growth of $PH$ on $NO_3$ & $\nu_{1}=\mu_{max,PH} \frac{S_{IC}}{K_{PH,IC}+S_{IC}} \frac{S_{NO_3}}{K_{PH,NO_3}+S_{NO_3}} \frac{K_{PH,NH_3}}{K_{PH,NH_3}+S_{NH_3}} \frac{K^{in}_{PH,O_2}}{K^{in}_{PH,O_2}+S_{O_2}} \frac{K^{in}_{M}}{K^{in}_{M}+M} \frac{I}{I_{opt}} \ e^{(1-(\frac{I}{I_{opt}})} \ f_{PH}$  \\

 \tiny{2} & Growth of $PH$ on $NH_3$ & $\nu_{2}=\mu_{max,PH} \frac{S_{IC}}{K_{PH,IC}+S_{IC}} \frac{S_{NH_3}}{K_{PH,NH_3}+S_{NH_3}} \frac{K^{in}_{PH,O_2}}{K^{in}_{PH,O_2}+S_{O_2}} \frac{K^{in}_{M}}{K^{in}_{M}+M} \frac{I}{I_{opt}} \ e^{(1-(\frac{I}{I_{opt}})} \ f_{PH}$ \\
  
 \tiny{3} & Heterotrophic growth of $PH$& $\nu_{3}= \mu^{resp}_{max,PH} \frac{S_{DOC}}{K_{PH,DOC}+S_{DOC}} \frac{S_{O_2}}{K_{PH,O_2}+S_{O_2}} \frac{K^{in}_{PH,I}}{K^{in}_{PH,I}+I} \frac{K^{in}_{M}}{K^{in}_{M}+M} \ f_{PH}$  \\
             
 \tiny{4} & Aerobic growth of $H$  & $\nu_{4}=\mu_{max,H} \frac{S_{DOC}}{K_{H,DOC}+S_{DOC}} \frac{S_{NH_3}}{K_{H,NH_3}+S_{NH_3}} \frac{S_{O_2}}{K_{H,O_2}+S_{O_2}}\frac{K^{in}_{M}}{K^{in}_{M}+M} \ f_H$  \\
  
 \tiny{5} & Anoxic growth of $H$  & $\nu_{5}=\mu_{max,H} \frac{S_{DOC}}{K_{H,DOC}+S_{DOC}} \frac{S_{NO_3}}{K_{H,NO_3}+S_{NO_3}} \frac{S_{NH_3}}{K_{H,NH_3}+S_{NH_3}}  \frac{K_{H,O_2}}{K_{H,O_2}+S_{O_2}} \frac{K^{in}_{M}}{K^{in}_{M}+M} \ f_H$  \\
              
 \tiny{6} & Growth of $N$  & $\nu_{6}=\mu_{max,N} \frac{S_{IC}}{K_{N,IC}+S_{IC}} \frac{S_{NH_3}}{K_{H,NH_3}+S_{NH_3}} \frac{S_{O_2}}{K_{H,O_2}+S_{O_2}} \frac{K^{in}_{M}}{K^{in}_{M}+M} \ f_N$  \\
 
 \tiny{7} & Death of $PH$  & $\nu_{7}= k_{d,PH}f_{PH}$ \\
  
 \tiny{8} & Death of $H$ & $\nu_{8}= k_{d,H}f_H$ \\
 
 \tiny{9} & Death of $N$ & $\nu_{9}= k_{d,N}f_N$ \\  
 
  \hline
  \multicolumn{3}{l}{where  $K^{in}_{PH,O_2}=K^{in}_{O_2,max} \ \frac{\frac{S_{IC}}{S_{O_2}}}{\frac{S_{IC}}{S_{O_2}}+K_{R_{IC/O_2}}}$} \\
  
        \end{tabular}
        \caption{Kinetic rate expressions ($\nu_{k}$) of the biological processes within the biofilm.}
        \label{t5.3}
         \end{center}
    \end{minipage}
    \vfill
    
        \begin{minipage}[b]{1\vsize}
        \begin{center}      
        
        \end{center}
    \end{minipage}%
}}%
}
\end{footnotesize}
\end{table}   
\clearpage

\clearpage
\begin{table}[ht!]
    \begin{center}  
        \begin{tabular}{|c|c|c|c|c|c|c|c|}
\hline
{\textbf{}} & {$\bf{\psi^*_{PH}}$} & {$\bf{\psi^*_{H}}$} & {$\bf{\psi^*_{N}}$} & {$\bf{\psi^*_{d_{PH}}}$} & {$\bf{\psi^*_{d_H}}$} & {$\bf{\psi^*_{d_N}}$} & {\textbf{Rate}} \\
 \hline
 \tiny{1} & \tiny{$1$} & & & \tiny{} & & & \tiny{$\nu^*_{1}$} \\
 \hline
 \tiny{2} & \tiny{$1$} & & & \tiny{} & & & \tiny{$\nu^*_{2}$} \\
 \hline
 \tiny{3} & \tiny{$1$} & & & \tiny{} & & & \tiny{$\nu^*_{3}$} \\
 \hline
 \tiny{4} & & \tiny{$1$} & & & \tiny{} & & \tiny{$\nu^*_{4}$} \\
 \hline
 \tiny{5} & & \tiny{$1$} & & & \tiny{} & & \tiny{$\nu^*_{5}$} \\
 \hline
 \tiny{6} & & & \tiny{$1$} & & & \tiny{} & \tiny{$\nu^*_{6}$} \\
 \hline
  \tiny{7} & \tiny{} & & & \tiny{$1$} & & & \tiny{$\nu^*_{7}$} \\
 \hline
 \tiny{8} & \tiny{} & & & \tiny{$1$} & & & \tiny{$\nu^*_{8}$} \\
 \hline
 \tiny{9} & \tiny{} & & & \tiny{$1$} & & & \tiny{$\nu^*_{9}$} \\
 \hline
 \tiny{10} & & \tiny{} & & & \tiny{$1$} & & \tiny{$\nu^*_{10}$} \\
 \hline
 \tiny{11} & & \tiny{} & & & \tiny{$1$} & & \tiny{$\nu^*_{11}$} \\
 \hline
 \tiny{12} & & & \tiny{} & & & \tiny{$1$} & \tiny{$\nu^*_{12}$} \\
 \hline
 \tiny{13} & \tiny{-1} & & & \tiny{} & & & \tiny{$\nu^*_{13}$} \\
 \hline
 \tiny{14} & & \tiny{-1} & & & \tiny{} & & \tiny{$\nu^*_{14}$} \\
 \hline 
 \tiny{15} & & & \tiny{-1} & & & \tiny{} & \tiny{$\nu^*_{15}$} \\
 \hline
 \tiny{16} & \tiny{} & & & \tiny{-1} & & & \tiny{$\nu^*_{16}$} \\
 \hline
 \tiny{17} & & \tiny{} & & & \tiny{-1} & & \tiny{$\nu^*_{17}$} \\
 \hline 
 \tiny{18} & & & \tiny{} & & & \tiny{-1} & \tiny{$\nu^*_{18}$} \\
 \hline
        \end{tabular}
        \caption{Biochemical rate coefficients ($\alpha^*_{i,k}$ and $\bar{\alpha}^*_{i,k}$) of the biological processes within the bulk liquid.}
        \label{t5.4}
         \end{center}

\begin{center}      
        \begin{tabular}{|c|c|c|c|c|c|c|}
 \hline
{\textbf{}} & {$\bf{S^*_{IC}}$} & {$\bf{S^*_{DOC}}$} & {$\bf{S^*_{NO_3}}$} & {$\bf{S^*_{NH_3}}$} & {$\bf{S^*_{O_2}}$} & {\textbf{Rate}}
 \\
 \hline
 \tiny{1} & \tiny{$-\frac{k_{DOC}+1.0025}{32}$} & \tiny{$k_{DOC}$} & \tiny{$-\frac{0.1704}{32}$} & & \tiny{$\frac{k_{DOC}+1}{32}$} & \tiny{$\nu^*_{1}$} \\
 \hline
 \tiny{2} & \tiny{$-\frac{k_{DOC}+1.0025}{32}$} & \tiny{$k_{DOC}$} & & \tiny{$-\frac{0.1704}{32}$} & \tiny{$\frac{k_{DOC}+1.3409}{32}$} & \tiny{$\nu^*_{2}$} \\
 \hline
 \tiny{3} & \tiny{$\frac{1-1.0025 \ Y_{DOC}}{32 \ Y_{DOC}}$} & \tiny{$-\frac{1}{Y_{DOC}}$} & & \tiny{$-\frac{0.1704}{32 \ Y_{DOC}}$} & \tiny{$-\frac{1-Y_{DOC}}{32 \ Y_{DOC}}$} & \tiny{$\nu^*_{3}$}\\
 \hline
 \tiny{4} & \tiny{$\frac{1}{32 \ Y_H}-0.02976$} & \tiny{$-\frac{1}{Y_H}$} & & \tiny{$-\frac{0.2}{33.6}$} & \tiny{$-\frac{1}{32 \ Y_H}+0.03125$} & \tiny{$\nu^*_{4}$} \\
 \hline
 \tiny{5} & \tiny{$\frac{1}{32 \ Y_H}-0.02976$} & \tiny{$-\frac{1}{Y_H}$} & \tiny{$-\frac{0.8}{32 \ Y_H}+0.02857$}  & \tiny{$-\frac{0.2}{33.6}$} &  & \tiny{$\nu^*_{5}$} \\
 \hline
 \tiny{6} & \tiny{$-\frac{1}{33.6}$} & & \tiny{$\frac{1}{14 \ Y_N}$} & \tiny{$-(0.00593 + \frac{1}{14 \ Y_N})$} & \tiny{$-\frac{1}{7 \ Y_N}+\frac{1}{32}$} & \tiny{$\nu^*_{6}$} \\
 \hline
 \tiny{7} & \tiny{$-\frac{k_{DOC}+1.0025}{32}$} & \tiny{$k_{DOC}$} & \tiny{$-\frac{0.1704}{32}$} & & \tiny{$\frac{k_{DOC}+1}{32}$} & \tiny{$\nu^*_{7}$} \\
 \hline
 \tiny{8} & \tiny{$-\frac{k_{DOC}+1.0025}{32}$} & \tiny{$k_{DOC}$} & & \tiny{$-\frac{0.1704}{32}$} & \tiny{$\frac{k_{DOC}+1.3409}{32}$} & \tiny{$\nu^*_{8}$} \\
 \hline
 \tiny{9} & \tiny{$\frac{1-1.0025 \ Y_{DOC}}{32 \ Y_{DOC}}$} & \tiny{$-\frac{1}{Y_{DOC}}$} & & \tiny{$-\frac{0.1704}{32 \ Y_{DOC}}$} & \tiny{$-\frac{1-Y_{DOC}}{32 \ Y_{DOC}}$} & \tiny{$\nu^*_{9}$}\\
 \hline
 \tiny{10} & \tiny{$\frac{1}{32 \ Y_H}-0.02976$} & \tiny{$-\frac{1}{Y_H}$} & & \tiny{$-\frac{0.2}{33.6}$} & \tiny{$-\frac{1}{32 \ Y_H}+0.03125$} & \tiny{$\nu^*_{10}$} \\
 \hline
 \tiny{11} & \tiny{$\frac{1}{32 \ Y_H}-0.02976$} & \tiny{$-\frac{1}{Y_H}$} & \tiny{$-\frac{0.8}{32 \ Y_H}+0.02857$}  & \tiny{$-\frac{0.2}{33.6}$} &  & \tiny{$\nu^*_{11}$} \\
 \hline
 \tiny{12} & \tiny{$-\frac{1}{33.6}$} & & \tiny{$\frac{1}{14 \ Y_N}$} & \tiny{$-(0.00593 + \frac{1}{14 \ Y_N})$} & \tiny{$-\frac{1}{7 \ Y_N}+\frac{1}{32}$} & \tiny{$\nu^*_{12}$} \\
 \hline
 \tiny{13} & & & & & & \tiny{$\nu^*_{13}$} \\
  \hline
 \tiny{14} & & & & & & \tiny{$\nu^*_{14}$} \\
  \hline
 \tiny{15} & & & & & & \tiny{$\nu^*_{15}$} \\
  \hline
 \tiny{16} & & & & & & \tiny{$\nu^*_{16}$} \\
  \hline
 \tiny{17} & & & & & & \tiny{$\nu^*_{17}$} \\
  \hline
 \tiny{18} & & & & & & \tiny{$\nu^*_{18}$} \\
 \hline
        \end{tabular}
        \caption{Biochemical rate coefficients ($\beta^*_{i,k}$) of the biological processes within the bulk liquid.}
        \label{t5.5}
        \end{center}
\end{table}  
\clearpage

\clearpage
\begin{table}[ht!]
\begin{footnotesize}
\centering
\rotatebox{90}{%
\makebox[0pt][c]{\parbox{3.0\textwidth}{%
    \begin{minipage}[b]{1\vsize}
    \begin{center}  
    \vspace{10mm}
        \begin{tabular}{|l|l|l|}
 \hline
{\textbf{}} & {\textbf{Process}} & {\textbf{Kinetic rate expression}} 
 \\
 \hline
 \tiny{1} & Growth of $PH$ on $NO_3$ & $\nu^*_{1}=\mu_{max,PH} \frac{S^*_{IC}}{K_{PH,IC}+S^*_{IC}} \frac{S^*_{NO_3}}{K_{PH,NO_3}+S^*_{NO_3}} \frac{K_{PH,NH_3}}{K_{PH,NH_3}+S^*_{NH_3}} \frac{K^{in,*}_{PH,O_2}}{K^{in,*}_{PH,O_2}+S^*_{O_2}} \frac{K^{in}_{M^*}}{K^{in}_{M}+M^*} \frac{I_0}{I_{opt}} \ e^{(1-\frac{I_0}{I_{opt}})}\psi^*_{PH}$  \\
 
  \tiny{2} & Growth of $PH$ on $NH_3$ & $\nu^*_{2}=\mu_{max,PH} \frac{S^*_{IC}}{K_{PH,IC}+S^*_{IC}} \frac{S^*_{NH_3}}{K_{PH,NH_3}+S^*_{NH_3}} \frac{K^{in,*}_{PH,O_2}}{K^{in,*}_{PH,O_2}+S^*_{O_2}} \frac{K^{in}_{M^*}}{K^{in}_{M}+M^*} \frac{I_0}{I_{opt,PH}} \ e^{(1-\frac{I_0}{I_{opt}})}\psi^*_{PH}$  \\
 
 \tiny{3} & Heterotrophic growth of $PH$ & $\nu^*_{3} =  \mu^{resp}_{max,PH} \frac{S^*_{DOC}}{K_{PH,DOC}+S^*_{DOC}} \frac{S^*_{O_2}}{K_{PH,O_2}+S^*_{O_2}} \frac{K^{in}_{PH,I}}{K^{in}_{PH,I}+I_0} \frac{K^{in}_{M^*}}{K^{in}_{M}+M^*} \psi^*_{PH}$  \\
      
 \tiny{4} & Aerobic growth of $H$  & $\nu^*_{4}=\mu_{max,H} \frac{S^*_{DOC}}{K_{H,DOC}+S^*_{DOC}} \frac{S^*_{NH_3}}{K_{H,NH_3}+S^*_{NH_3}} \frac{S^*_{O_2}}{K_{H,O_2}+S^*_{O_2}}\frac{K^{in}_{M^*}}{K^{in}_{M}+M^*} \psi^*_{H}$  \\
  
 \tiny{5} & Anoxic growth of $H$  & $\nu^*_{5}=\mu_{max,H} \frac{S^*_{DOC}}{K_{H,DOC}+S^*_{DOC}} \frac{S^*_{NO_3}}{K_{H,NO_3}+S^*_{NO_3}} \frac{S^*_{NH_3}}{K_{H,NH_3}+S^*_{NH_3}} \frac{K_{H,O_2}}{K_{H,O_2}+S^*_{O_2}} \frac{K^{in}_{M^*}}{K^{in}_{M}+M^*} \psi^*_{H}$  \\
              
 \tiny{6} & Growth of $N$  & $\nu_{6}=\mu^*_{max,N} \frac{S^*_{IC}}{K_{N,IC}+S^*_{IC}} \frac{S^*_{NH_3}}{K_{H,NH_3}+S^*_{NH_3}} \frac{S^*_{O_2}}{K_{H,O_2}+S^*_{O_2}} \frac{K^{in}_{M^*}}{K^{in}_{M}+M^*}\psi^*_{N}$  \\
 
  \tiny{7} & Growth of $PH_d$ on $NO_3$ & $\nu^*_{7}=\mu_{max,PH} \frac{S^*_{IC}}{K_{PH,IC}+S^*_{IC}} \frac{S^*_{NO_3}}{K_{PH,NO_3}+S^*_{NO_3}} \frac{K_{PH,NH_3}}{K_{PH,NH_3}+S^*_{NH_3}} \frac{K^{in,*}_{PH,O_2}}{K^{in,*}_{PH,O_2}+S^*_{O_2}} \frac{K^{in}_{M^*}}{K^{in}_{M}+M^*} \frac{I_0}{I_{opt}} \ e^{(1-\frac{I_0}{I_{opt}})} \ \psi^*_{d_{PH}}$  \\
 
  \tiny{8} & Growth of $PH_d$ on $NH_3$ & $\nu^*_{8}=\mu_{max,PH} \frac{S^*_{IC}}{K_{PH,IC}+S^*_{IC}} \frac{S^*_{NH_3}}{K_{PH,NH_3}+S^*_{NH_3}} \frac{K^{in,*}_{PH,O_2}}{K^{in,*}_{PH,O_2}+S^*_{O_2}} \frac{K^{in}_{M^*}}{K^{in}_{M}+M^*} \frac{I_0}{I_{opt,PH}} \ e^{(1-\frac{I_0}{I_{opt}})}\psi^*_{d_{PH}}$  \\
 
 \tiny{9} & Heterotrophic growth of $PH_d$ & $\nu^*_{9} = \mu^{resp}_{max,PH} \frac{S^*_{DOC}}{K_{PH,DOC}+S^*_{DOC}} \frac{S^*_{O_2}}{K_{PH,O_2}+S^*_{O_2}} \frac{K^{in}_{PH,I}}{K^{in}_{PH,I}+I_0} \frac{K^{in}_{M^*}}{K^{in}_{M}+M^*} \psi^*_{d_{PH}}$  \\
      
 \tiny{10} & Aerobic growth of $H_d$  & $\nu^*_{10}=\mu_{max,H} \frac{S^*_{DOC}}{K_{H,DOC}+S^*_{DOC}} \frac{S^*_{NH_3}}{K_{H,NH_3}+S^*_{NH_3}} \frac{S^*_{O_2}}{K_{H,O_2}+S^*_{O_2}}\frac{K^{in}_{M^*}}{K^{in}_{M}+M^*} \psi^*_{d_H}$  \\
  
 \tiny{11} & Anoxic growth of $H_d$  & $\nu^*_{11}=\mu_{max,H} \frac{S^*_{DOC}}{K_{H,DOC}+S^*_{DOC}} \frac{S^*_{NO_3}}{K_{H,NO_3}+S^*_{NO_3}} \frac{S^*_{NH_3}}{K_{H,NH_3}+S^*_{NH_3}} \frac{K_{H,O_2}}{K_{H,O_2}+S^*_{O_2}} \frac{K^{in}_{M^*}}{K^{in}_{M}+M^*}\psi^*_{d_H}$  \\
              
 \tiny{12} & Growth of $N_d$  & $\nu^*_{12}=\mu^*_{max,N} \frac{S^*_{IC}}{K_{N,IC}+S^*_{IC}} \frac{S^*_{NH_3}}{K_{H,NH_3}+S^*_{NH_3}} \frac{S^*_{O_2}}{K_{H,O_2}+S^*_{O_2}} \frac{K^{in}_{M^*}}{K^{in}_{M}+M^*}\psi^*_{d_N}$  \\
 
 \tiny{13} & Death of $PH$  & $\nu^*_{13}= k_{d,PH}\psi^*_{PH}$ \\
 
 \tiny{14} & Death of $H$ & $\nu^*_{14}= k_{d,H}\psi^*_{H}$ \\
 
 \tiny{15} & Death of $N$ & $\nu^*_{15}= k_{d,N}\psi^*_{N}$ \\
 
  \tiny{16} & Death of $PH_d$  & $\nu^*_{16}= k_{d,PH}\psi^*_{d_{PH}}$ \\
 
 \tiny{17} & Death of $H_d$ & $\nu^*_{17}= k_{d,H}+\psi^*_{d_H}$ \\
 
 \tiny{18} & Death of $N_d$ & $\nu^*_{18}= k_{d,N}\psi^*_{d_N}$ \\  
   
 \hline
  \multicolumn{3}{l}{where $K^{in,*}_{PH,O_2}=K^{in}_{O_2,max} \ \frac{\frac{S^*_{IC}}{S^*_{O_2}}}{\frac{S^*_{IC}}{S^*_{O_2}}+K_{R_{IC/O_2}}}$}\\
        \end{tabular}
        \caption{Kinetic rate expressions ($\nu^*_{k}$) of the biological processes within the bulk liquid.}
        \label{t5.6}
         \end{center}
    \end{minipage}
    \vfill
    
        \begin{minipage}[b]{1\vsize}
        \begin{center}      
        
        \end{center}
    \end{minipage}%
}}%
}
\end{footnotesize}
\end{table}   
\clearpage

\clearpage
\begin{table}[ht!]
\begin{scriptsize}
\centering
\begin{tabular}{llccc}
 \hline
{\textbf{Parameter}} & {\textbf{Definition}} & {\textbf{Unit}} & {\textbf{Value}} & {\textbf{Ref}}
 \\
 \hline
 $\mu_{max,PH}$ & Maximum specific growth rate for $PH$ &  $d^{-1}$ & $2.368$ & \cite{wolf2007kinetic} \\
  $\mu^{resp}_{max,PH}$ & Maximum specific growth rate for $PH$ respiration &  $d^{-1}$ & $0.237$ & \cite{wolf2007kinetic} \\
 $\mu_{max,H}$ & Maximum specific growth rate for $H$ &  $d^{-1}$ & $4.8$  & \cite{munoz2014modeling}\\
  $\mu_{max,N}$ & Maximum specific growth rate for $N$ &  $d^{-1}$  & $1$ & \cite{munoz2014modeling} \\
 $k_{d,PH}$ & Decay-inactivation rate for $PH$  &  $d^{-1}$ & $0.1$  & \cite{wolf2007kinetic} \\
 $k_{d,H}$  & Decay-inactivation rate for $H$ &  $d^{-1}$ & $0.1$  & \cite{wolf2007kinetic} \\
 $k_{d,N}$ & Decay-inactivation rate for $N$ &  $d^{-1}$  & $0.1$  & \cite{wolf2007kinetic} \\
 $K_{PH,IC}$    & $IC$ half saturation coeff. for $PH$ &  $kmol(IC) \ m^{-3}$ & $10^{-4}$ & \cite{wolf2007kinetic} \\
  $K_{PH,DOC}$ & $DOC$ half saturation coeff. for $PH$ & $kg(COD) \ m^{-3}$ & $5 \cdot10^{-3}$ & \cite{wolf2007kinetic} \\
 $K_{PH,NO_3}$ & $NO_3$ half saturation coeff. for $PH$  &  $kmol(NO_3) \ m^{-3}$ & $1.2\cdot10^{-6}$ & \cite{wolf2007kinetic} \\
 $K_{PH,NH_3}$ & $NH_3$ half saturation coeff. for $PH$ &  $kmol(NH_3) \ m^{-3}$ & $1.2\cdot10^{-6}$ &\cite{wolf2007kinetic} \\
  $K_{PH,O_2}$ & $O_2$ half saturation coeff. for $PH$ &  $kmol(O_2) \ m^{-3}$ & $3\cdot10^{-4}$ &\cite{wolf2007kinetic} \\
 $K^{in}_{PH,I}$ & Light inhibition coefficient for $PH$ & $kmol(e^-) \ m^{-2} \ d^{-1}$ & $8\cdot10^{-5}$ & \cite{wolf2007kinetic} \\
 $K_{H,DOC}$ & $DOC$ half saturation coeff. for $H$ & $kg(COD) \ m^{-3}$ & $4 \cdot10^{-3}$ & \cite{wolf2007kinetic} \\
 $K_{H,NO_3}$ & $NO_3$ half saturation coeff. for $H$  &  $kmol(NO_3) \ m^{-3}$ & $3.6\cdot10^{-5}$ & \cite{wolf2007kinetic} \\
 $K_{H,NH_3}$ & $NH_3$ half saturation coeff. for $H$ &  $kmol(NH_3) \ m^{-3}$ & $3.6\cdot10^{-6}$ & \cite{henze2000activated} \\
  $K_{H,O_2}$ & $O_2$ half saturation coeff. for $H$ & $kmol(O_2) \ m^{-3}$ & $6.25\cdot10^{-6}$ & \cite{wolf2007kinetic} \\
 $K_{N,IC}$    & $IC$ half saturation coeff. for $N$ &  $kmol(IC) \ m^{-3}$ & $10^{-4}$ & \cite{wolf2007kinetic} \\
 $K_{N,NH_3}$ & $NH_3$ half saturation coeff. for $N$ &  $kmol(NH_3) \ m^{-3}$ & $7\cdot10^{-5}$ &\cite{wolf2007kinetic} \\
  $K_{N,O_2}$ & $O_2$ half saturation coeff. for $N$ & $kmol(O_2) \ m^{-3}$ & $1.56\cdot10^{-5}$ & \cite{wolf2007kinetic} \\
  $K^{in}_{O_2,max}$ & Max inhibition coefficient of $O_2$ on $PH$ &   $kmol(O_2) \ m^{-3}$ & $10^{-3}$     &   \cite{li2016investigating}    \\ 
  $K_{R_{IC/O_2}}$ & Half saturation coeff. for $O_2$ inhibition     &  $--$ & $0.35$     &  \cite{li2016investigating}    \\ 
 $K^{in}_{M}$ & Inhibition coefficient of $M$  & $kg(M) \ m^{-3}$ & $0.1$ & (a)    \\ 
 $Y_{H}$ & Yield of $H$ on $DOC$  & $kg(COD) \ kg(COD)^{-1}$ & $0.63$ & \cite{wolf2007kinetic} \\ 
 $Y_{N}$ & Yield of $N$ on $NO_3$  & $kg(COD) \ kg(NO_3-N)^{-1}$ & $0.24$ & \cite{wolf2007kinetic}  \\
 $Y_{DOC}$ & Yield of $PH$ on $DOC$  & $kg(COD) \ kg(COD)^{-1}$ & $0.5$ & (a) \\  
 $\tilde{k}_{EPS,PH}$ & EPS fraction produced by $PH$  & $--$ & $0.23$ & (a)  \\
 $\tilde{k}_{EPS,H}$ & EPS fraction produced by $H$  & $--$ & $0.18$ & \cite{merkey2009modeling}  \\
 $\tilde{k}_{EPS,N}$ & EPS fraction produced by $N$  & $--$ & $0.075$ & \cite{merkey2009modeling}  \\
 $k_{DOC}$ & DOC release fraction by $PH$  & $--$ & $0.05$ & \cite{tenore2021modelling}\\
 $k_{La}$ & $O_2$ mass transfer coefficient & $d^{-1}$ & $23.3$ & \cite{munoz2014modeling} \\
  $S_{O_2,sat}$ & $O_2$ saturation concentration in bulk liquid & $kmol(O_2) \ m^{-3}$ & $2.4\cdot10^{-4}$ & \cite{munoz2014modeling} \\
 $I_{opt}$ & Optimum light intensity for $PH$                   & $kmol (e^-) \ m^{-2} \ d^{-1}$ & $0.01728$   &   \cite{flora1995modeling}   \\
  $I_{0}$ & Incident light intensity in the reactor              & $kmol (e^-) \ m^{-2} \ d^{-1}$ & $0.008$ & \cite{tenore2021multiscale}   \\
 $k_{tot}$ & Light attenuation coefficient & $m^{2} \ kg^{-1}$ & $210$   &   \cite{wolf2007kinetic}  \\
  $k_{ads,PH}$  & Sorption constant of $PH$ & $m^3 \ kg(M)^{-1} \ d^{-1} $ & $1\cdot10^{3}$ & (a)  \\
  $k_{ads,H}$  & Sorption constant of $H$ & $m^3 \ kg(M)^{-1} \ d^{-1} $ & $2\cdot10^{1}$ & (a) \\
  $k_{ads,N}$  & Sorption constant of $N$ & $m^3 \ kg(M)^{-1} \ d^{-1} $ & $2\cdot10^{1}$ & (a) \\
  $k_{ads,EPS}$  & Sorption constant of $EPS$ & $m^3 \ kg(M)^{-1} \ d^{-1} $ & $2\cdot10^{3}$ & (a) \\
  $k_{ads,I}$  & Sorption constant of $I$ & $m^3 \ kg(M)^{-1} \ d^{-1} $ & $2\cdot10^{2}$ & (a) \\
  ${K}_{s,i}$  & Stimulation constant for $EPS$ & $kg(M) \ m^{-3}$ & $0.05$ & (a) \\
  $Y_{ads,i}$  & Yield of $M$ on $i^{th}$ microbial species & $kg(M) \ N^{{\circ}^{-1}}_{sites}$ & 1 & (a) \\
 $D_{S,IC}$ & Diffusion coefficient of $IC$ in biofilm & 
 $m^2 \ d^{-1}$ & $1.32\cdot10^{-4}$ & \cite{wolf2007kinetic} \\
 $D_{S,DOC}$ & Diffusion coefficient of $DOC$ in biofilm & $m^2 \ d^{-1}$ & $0.83\cdot10^{-4}$ & \cite{wanner1986multispecies} \\
 $D_{S,NO_3}$ & Diffusion coefficient of $NO_3$ in biofilm & $m^2 \ d^{-1}$ & $1.18\cdot10^{-4}$ &  \cite{wolf2007kinetic} \\
 $D_{S,NH_3}$ & Diffusion coefficient of $NH_3$ in biofilm & $m^2 \ d^{-1}$ & $1.49\cdot10^{-4}$ & \cite{wanner1986multispecies} \\
 $D_{S,O_2}$ & Diffusion coefficient of $O_2$ in biofilm & $m^2 \ d^{-1}$ & $1.75\cdot10^{-4}$ & \cite{wanner1986multispecies} \\
  $D_{S,M}$ & Diffusion coefficient of metal in biofilm & $m^2 \ d^{-1}$ & $6.05\cdot10^{-5}$ & \cite{d2016mathematical}  \\
   $v^0_{a,PH}$ & Attachment velocity of $\psi^*_{PH}$ & $m \ d^{-1}$ & $3\cdot10^{-3}$    &    (a)    \\
   $v^0_{a,H}$ & Attachment velocity of $\psi^*_{H}$ & $m \ d^{-1}$ & $5\cdot10^{-4}$    &    (a)    \\
   $v^0_{a,N}$ & Attachment velocity of $\psi^*_{N}$ & $m \ d^{-1}$  & $5\cdot10^{-4}$    &    (a)    \\
$K_{PH}$  & Half saturation coeff. of $\psi^*_{PH}$ on $\psi^*_{H}$, $\psi^*_{N}$ attachment & $kg (COD) \ m^{-3}$  & $3\cdot10^{-2}$ & (a) \\
$K_{C}$ & Conversion coeff. from detached to planktonic form & $d^{-1}$  & $0.5$ & (a)    \\
 $\rho$ & Biofilm density  & $kg(COD) \ m^{-3}$ & $37$  & \cite{munoz2014modeling}  \\
 $\rho_{\theta}$  & Binding sites density & $N^{\circ}_{sites} \ m^{-3}$ & $20$ & (a)  \\
 $\lambda$  & Constant detachment coefficient   &  $m^{-1} \ d^{-1}$ & $40$  &  (a)   \\
  $V$  & Reactor volume  &  $m^{3}$   & 400  &    (a)   \\
  $N_G$  &  Number of granules in the reactor &  $--$   & $2.4\cdot10^{10}$ & (a)   \\
  $\tau$  &  Duration of the cycle &  $d$   & $0.25$ & (a)   \\
  $\gamma$  &  Fraction of suspended biomass lost in the emptying &  $--$   & $0.2$ & (a)   \\
  $\omega$  &  Emptying/refilling ratio &  $--$   & $0.5$ & (a)   \\
$t_{light}$  &  Time of light condition &  $d$   & $0.125$ & (a)   \\
$t_{dark}$  &  Time of dark condition &  $d$   & $0.125$ & (a)   \\
 \hline
 \multicolumn{5}{l}{(a) Assumed}\\
\end{tabular}
\caption{Kinetic, stoichiometric and operating parameters used for numerical simulations.} \label{t5.7}
\end{scriptsize}
\end{table}
\clearpage

\section{Numerical studies and results} \label{n5.4}
\
 The model has been integrated numerically by developing an original code in MatLab platform. Hyperbolic PDEs \eqref{5.2.1.7} and \eqref{5.2.1.18} have been integrated by using the method of characteristics, applied for the first time in the planar biofilm context by D'Acunto and Frunzo (2011) \cite{d2011qualitative}. The method of lines has been used to solve the diffusion-reaction PDEs \eqref{5.2.1.14} and \eqref{5.2.1.19}. The ODEs for $\psi^*_{i}$, $\psi^*_{d_i}$, $S^*_{j}$, and $M^*_{j}$ (Eqs. \eqref{5.2.2.5}-\eqref{5.2.2.8}) have been integrated by using the MatLab routine ode45. Numerical simulations have been performed to investigate the genesis and evolution of oxygenic photogranules, the microbial species stratification and interaction between the functional trophic groups, and to study the SBR performances in terms of substrates removal and metal adsorption. Specifically, the first study ($SET1$) investigates the treatment process of a typical industrial wastewater containing a low concentration of a generic metal. Both the granules ecology and the process evolution have been investigated, focusing on metal effects on biofilm formation and its removal process. The second study ($SET2$) investigates how the metal concentration affects the OPGs formation and adsorption processes in terms of microbial growth and removal efficiency. Finally, the third study ($SET3$) explores the role of the adsorption capacities of all the microbial species in the adsorption process. 
 
The wastewater influent is supposed to be fed discontinuously in the SBR. As mentioned before, in each cycle the reactor is filled with a fixed volume $V$ of wastewater, and the substrates are biologically degraded in batch conditions. The bioreactor volume $V$ is assumed constant and equal to $400 \  m^3$. The number of granules $N_{G}$ has been selected through an iterative procedure varying the detachment coefficient $\lambda$ \cite{tenore2021multiscale1}, with the aim to obtain a $25$\% filling ratio by considering granules with a steady-state radius of about $1 \ mm$ (an average size representative of OPGs \cite{abouhend2019growth, abouhend2018oxygenic}). After the reaction phase, the solid-liquid separation occurs in the reactor, whereby perfect settling has been considered for granules (no granule is removed from the reactor during the emptying phase). While the fraction of suspended biomass lost during the emptying phase has been set equal to $20$\%. At the end of each cycle the reactor is only partially emptied and refilled with a new liquid volume to be treated (emptying/refilling ratio $\omega=50$\%). As explained above, the reaction phase is supposed to be the same as the duration of the cycle $\tau$, and it consists of $3$ hours of darkness and $3$ hours of light ($t_{dark}=t_{light}=0.125 \ d$ and $\tau=0.25 \ d$) \cite{tenore2021multiscale, abouhend2018oxygenic}. In the light phase, the reactor is supposed to be homogeneously illuminated and the incident light intensity $I_0$ is fixed at $0.008 \ kmol \ m^{-2} \ d^{-1}$ \cite{tenore2021multiscale}. 

 The same wastewater influent composition is considered for each treatment cycle. It is characterized by $S^{in}_{IC} = 180 \ g \  m^{-3}$ (inorganic carbon), $S^{in}_{DOC} = 500 \ g \  m^{-3}$ (organic carbon), $S^{in}_{NH_3} = 50 \ g \  m^{-3}$ (ammonia), $S^{in}_{NO_3} = 0$ (nitrate), $S^{in}_{O_2} = 0$ (oxygen). Such concentrations reflect the typical wastewater \cite{brockmann2021wastewater} and are usually used in experimental works \cite{yang2020enhanced}. The concentration of the heavy metal in the influent $M^{in}$ is varied in the numerical studies, and its values will be provided below, case to case. The initial concentrations of soluble substrates $S^*_{j,0}$ and metal $M^*_{0}$ in the bulk liquid have been set equal to the concentration within the wastewater influent ($S^{in}_j$ and $M^{in}$). On the contrary, no suspended biomass is supposed to be present in the influent ($\psi^{in}_i=0$ and $\psi^{in}_{d_i}=0$), while phototrophic inoculum of suspended phototrophs is considered, where planktonic heterotrophic and nitrifying bacteria are present in smaller amounts: $\psi^*_{PH,0} = 600 \ g \  m^{-3}$, $\psi^*_{H,0} = \psi^*_{N,0} = 50 \ g \  m^{-3}$ \cite{tenore2021multiscale}. The initial concentration of detached species $\psi^*_{d_i,0}$ in the bulk liquid has been set equal to zero ($\psi^*_{d_i,0} =0$). Note that no addition of oxygen is considered ($S^{in}_{O_2}=0$), since it is provided by photosynthesis of phototrophs. 

Since the phototrophs are the major $EPS$ producers in algal-bacterial biofilm, their $EPS$ fraction produced in absence of toxic pollutants is assumed to be higher than heterotrophs ($\tilde{K}_{EPS,H} = 0.18$ \cite{merkey2009modeling}) and nitrifiers ($\tilde{K}_{EPS,N} = 0.075$ \cite{merkey2009modeling}) and is fixed at $\tilde{K}_{EPS,PH} = 0.23$. Such value is within the range of typical $EPS$ fraction values of phototrophic biomass \cite{wolf2007kinetic}. Regarding the adsorption kinetic constants of the biofilm components, $k_{ads,EPS}$, $k_{ads,PH}$ and $k_{ads,I}$ are supposed to be much higher than $k_{ads,H}$ and $k_{ads,N}$, since the adsorption process is predominantly governed by $EPS$, phototrophs and inactive material. Finally, as mentioned above, attachment velocity of phototrophs is assumed to be a constant value and it is set equal to the average value of attachment velocities of microalgae and cyanobacteria used by Tenore et al. \cite{tenore2021multiscale}. All parameters used in this model are reported in Table \ref{t5.7}.
  
 The simulation time $T$ is fixed to $200 \ d$ for all simulations. This time interval guarantees to achieve the steady-state configuration in terms of: performance of SBR cycles (including soluble substrates $S^*_{j}$, metal $M^*$, planktonic species $\psi^*_{i}$, and detached biomasses $\psi^*_{d_i}$), granule size $R(t)$; microbial composition and distribution within the granules (in terms of fraction $f_i$ and mass $m_i$); volume fractions of free binding sites $\theta_i$; and concentration of free metal $M$ within the biofilm. 
 
\subsection{SET1 - Evaluation of metal removal from industrial wastewater in OPGs-based system} \label{n5.4.1}

\begin{figure*}  
\centering
\fbox{\includegraphics[width=1\textwidth, keepaspectratio]{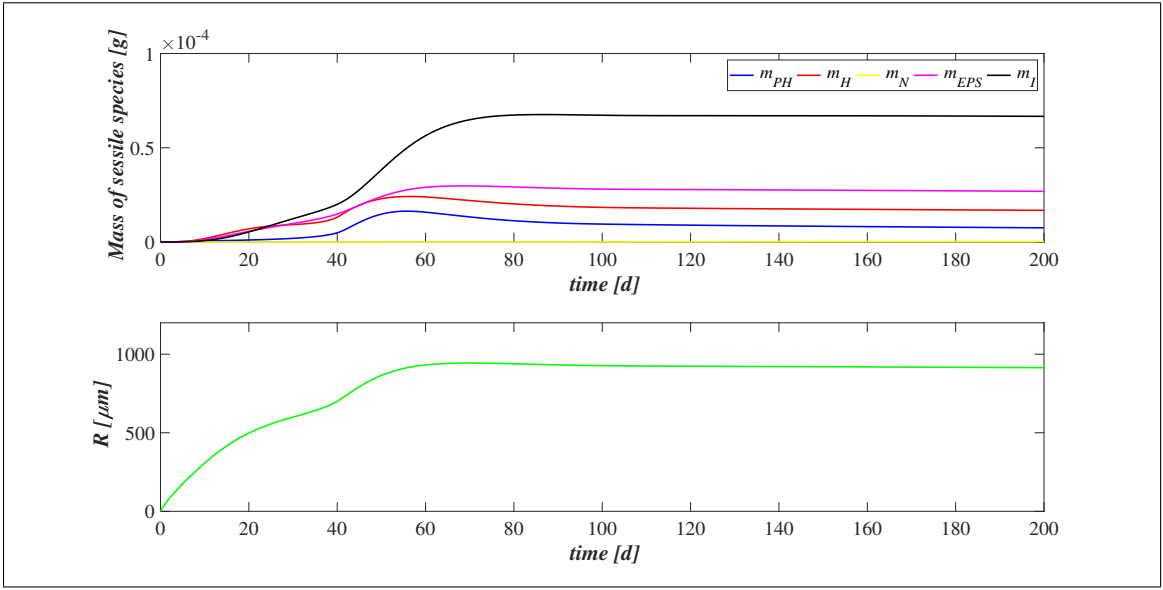}}
\caption{$SET1$ - Evolution of biofilm radius and mass of sessile species over time. Wastewater influent composition: $S^{in}_{IC} = 180 \ g \ m^{-3}$ (inorganic carbon), $S^{in}_{DOC} = 500 \ g \ m^{-3}$ (organic carbon), $S^{in}_{NH_3} = 50 \ g \ m^{-3}$ (ammonia), $S^{in}_{NO_3} = 0$ (nitrate), $S^{in}_{O_2} = 0$ (oxygen), $M^{in}=100 \ g \ m^{-3}$ (metal). Incident light intensity: $I_0= 0.008 \ kmol \ m^{-2} \ d^{-1}$. Duration of the cycle: $\tau =6 \ h$. Time of light exposure: $t_{light} = 50$\% $\tau$.} \label{f5.4.1.1}
\end{figure*}

\begin{figure*}    
\centering
\fbox{\includegraphics[width=1\textwidth, keepaspectratio]{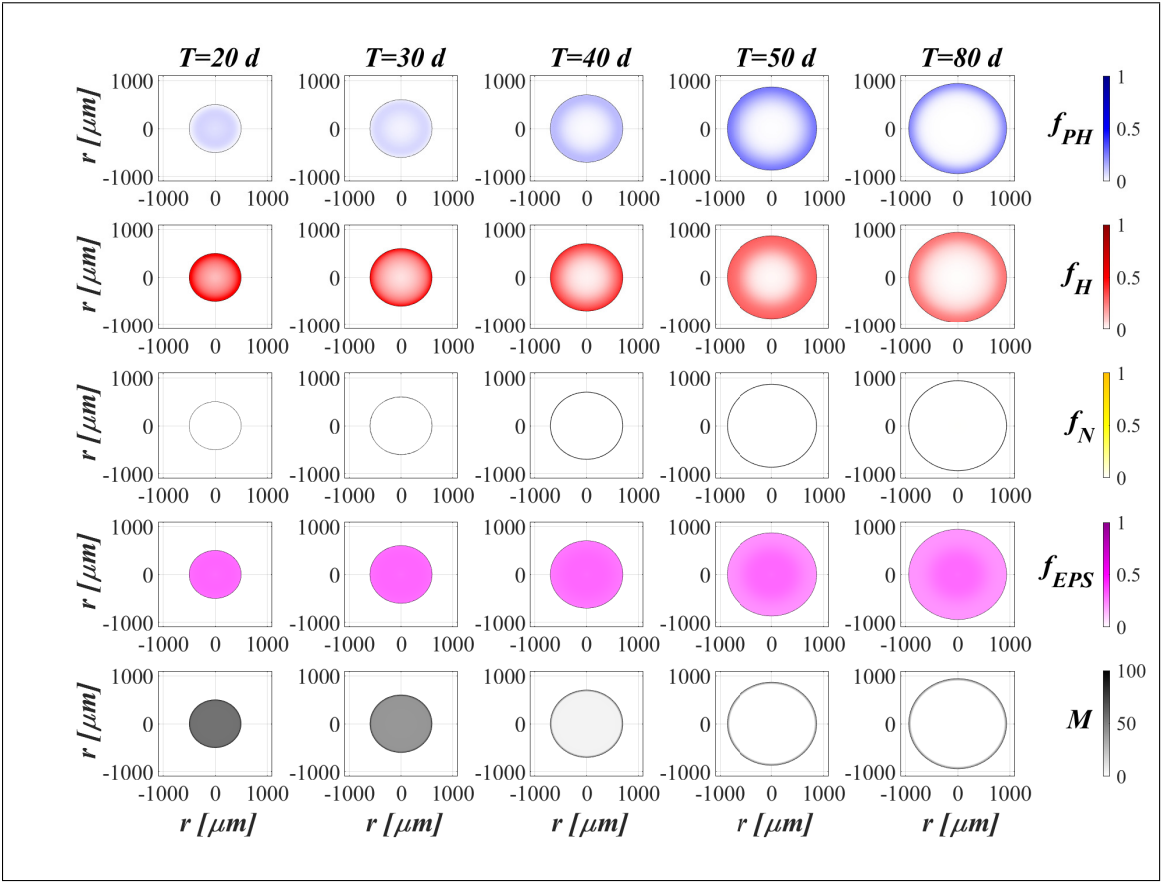}}
\caption{$SET1$- Active microbial species distribution and concentration of free metal within the diametrical section of the granule, at $T = 20 \ d$, $T = 30 \ d$, $T = 40 \ d$, $T = 50 \ d$, $T = 80 \ d$. Wastewater influent composition: $S^{in}_{IC} = 180 \ g \ m^{-3}$ (inorganic carbon), $S^{in}_{DOC} = 500 \ g \ m^{-3}$ (organic carbon), $S^{in}_{NH_3} = 50 \ g \ m^{-3}$ (ammonia), $S^{in}_{NO_3} = 0$  (nitrate), $S^{in}_{O_2}= 0$ (oxygen), $M^{in}=100 \ g \ m^{-3}$ (metal). Incident light intensity: $I_0= 0.008 \ kmol \ m^{-2} \ d^{-1}$. Duration of the cycle: $\tau =6 \ h$. Time of light exposure: $t_{light} = 50$\% $\tau$.} \label{f5.4.1.2}
\end{figure*}

\begin{figure*}
\centering
\fbox{\includegraphics[width=1\textwidth, keepaspectratio]{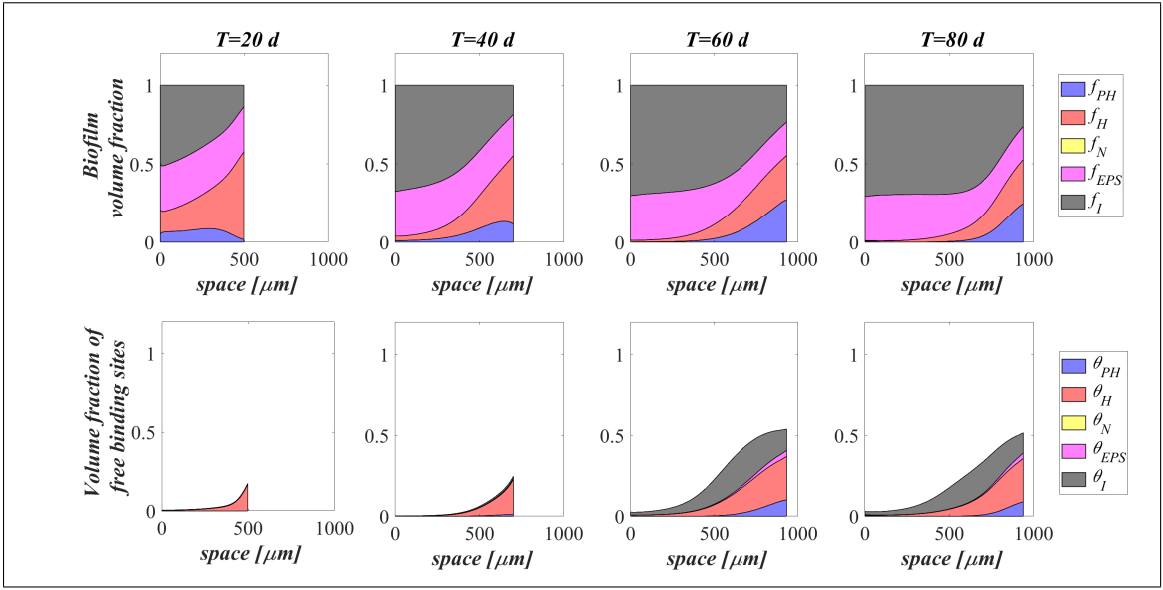}}
\caption{$SET1$ - Distribution of microbial species and free binding sites across the radius of the granule, at $T = 20 \ d$, $T = 40 \ d$, $T = 60 \ d$, $T = 80 \ d$. Wastewater influent composition: $S^{in}_{IC} = 180 \ g \ m^{-3}$ (inorganic carbon), $S^{in}_{DOC} = 500 \ g \ m^{-3}$ (organic carbon), $S^{in}_{NH_3} = 50 \ g \ m^{-3}$ (ammonia), $S^{in}_{NO_3} = 0$ (nitrate), $S^{in}_{O_2} = 0$ (oxygen), $M^{in}=100 \ g \ m^{-3}$ (metal). Incident light intensity: $I_0= 0.008 \ kmol \ m^{-2} \ d^{-1}$. Duration of the cycle: $\tau =6 \ h$. Time of light exposure: $t_{light} = 50$\% $\tau$.} \label{f5.4.1.3}
\end{figure*}

\begin{figure*}    
\centering
\fbox{\includegraphics[width=1\textwidth, keepaspectratio]{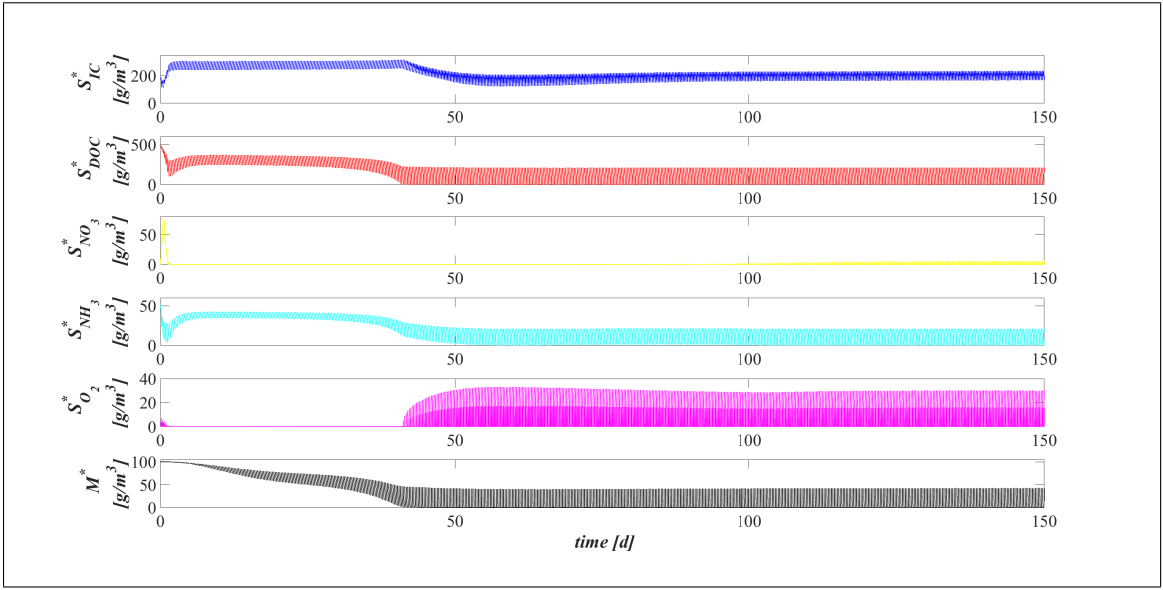}}
\caption{$SET1$ - Evolution of soluble substrates and metal concentrations within the reactor. Wastewater influent composition: $S^{in}_{IC} = 180 \ g \ m^{-3}$ (inorganic carbon), $S^{in}_{DOC} = 500 \ g \ m^{-3}$ (organic carbon), $S^{in}_{NH_3} = 50 \ g \ m^{-3}$ (ammonia), $S^{in}_{NO_3} = 0$ (nitrate), $S^{in}_{O_2} = 0$ (oxygen), $M^{in}=100 \ g \ m^{-3}$ (metal). Incident light intensity: $I_0= 0.008 \ kmol \ m^{-2} \ d^{-1}$. Duration of the cycle: $\tau =6 \ h$. Time of light exposure: $t_{light} = 50$\% $\tau$.} \label{f5.4.1.4}
\end{figure*}

The first set $SET1$ describes the treatment process of a typical industrial wastewater with a low concentration of metal ($M^{in} = 100 \ g \ m^{-3} $), occurring in a granular-based sequencing batch reactor (simulation $S1$). The microbial stratification of oxygenic photogranules, nutrients degradation, and metal adsorption are investigated. The $SET1$ results are shown in Figs. \ref{f5.4.1.1}-\ref{f5.4.1.4}.

 Fig. \ref{f5.4.1.1} reports the evolution of the overall mass of sessile species $m_i(t)$ (top) and photogranule radius $R(t)$ (bottom) over time. The active sessile biomasses constituting the biofilm matrix (phototrophs, heterotrophs and nitrifiers) grow by converting the nutrients, and decay producing inert material. Such biomasses can interact with each other, cooperating and/or competing. In presence of light, phototrophs produce $O_2$ and $DOC$ consuming $IC$ and $NH_3$ (or $NO_3$), and consequently they promote the heterotrophs and nitrifiers growth. Indeed, heterotrophic bacteria require $O_2$ (or $NO_3$ in anoxic condition) and $DOC$ for their metabolic activities. While $NH_3$ and $O_2$ are necessary for nitrifying bacteria, which compete with phototrophs for $IC$. However, nitrifiers produce $NO_3$ necessary for heterotrophs in anoxic condition and for phototrophs in lack or shortage of $NH_3$. Under dark conditions, phototrophs compete with heterotrophs and nitrifiers for $O_2$ and with all heterotrophs for $DOC$. In return, heterotrophic bacteria produce $IC$ necessary for the metabolism of phototrophs in light conditions and nitrifiers. In the initial days, the intense attachment process leads to the formation of photogranules mainly composed by phototrophs. Oxygen production during the photosynthesis promotes the growth of heterotrophic bacteria. Thus, the heterotrophic biomass rapidly increases with respect to the other microbial species, thanks to their high growth rates in presence of elevate availability of $DOC$. It should be noted that the metabolism of all biomasses is initially inhibited, due to the presence of free metal. In this phase the photogranule slowly increases, achieving a radius of about $ 600 \ \mu m$ (Fig.  \ref{f5.4.1.1} - bottom). After $40$ days, when a relevant amount of metal is already adsorbed on biofilm matrix, a more rapid phototrophs (blue) growth is observed. As consequence, heterotrophs (red) metabolism and production of $EPS$ (magenta) and inert material (black) are favoured. This, in turn, leads to a faster increase of the granule radius. Subsequently, the detachment process becomes more relevant and limits the granule expansion leading to a steady-state dimension of about $920 \ \mu m$. A very low mass of nitrifying bacteria (yellow) is observed throughout the process, because they have lower maximum growth rates than heterotrophic bacteria which are more competitive in the use of $O_2$ in presence of $DOC$. 

Fig. \ref{f5.4.1.2} shows the microbial stratification within the granule (from first row to fourth row) and the free metal concentration (fifth row) at different times. After $20$ and $30$ days, significant fractions of phototrophs (blue) and heterotrophs (red) can be observed throughout the granule. Indeed, phototrophs are responsible for the genesis of the photogranules due to their granulation properties, while heterotrophic bacteria have the highest growth rate. Obviously, in the initial stage of the process the concentration of free metal (black) is still elevate and the metal diffuses throughout the granule inhibiting the microorganisms growth. Passing from $T=30 \ d$ to $T=40 \ d$ the concentration of free metal significantly reduces thanks to the growth of phototrophs (see Fig. \ref{f5.4.1.4}). Indeed, phototrophs have a tendency to secrete $EPS$ higher than other microbial species, and both phototrophs and $EPS$ have higher adsorption capabilities than heterotrophs and nitrifiers. Thanks to the metal consumption, phototrophs are in turn less inhibited and continue to grow. The steady-state of microbial species distribution and free metal concentration is achieved at $T=80 \ d$. Note that phototrophs and $EPS$ are present in relevant amounts and a clear microbial species stratification can be observed: phototrophic biomass accumulates in the outermost layers, where optimal light conditions are guaranteed; heterotrophic bacteria predominantly populate the external part of the granule; $EPS$ (magenta) is homogeneously distributed throughout the granule. In addition, as observed in Fig. \ref{f5.4.1.1}, nitrifying bacteria (yellow) are almost absent. As regard the free metal diffusion, after the complete evolution of the granule, the adsorption process is completed, and a gradient of free metal concentration can be observed across the granule: the free metal concentration goes from low values in a thin external layer to zero in the internal part. 

Biofilm volume fractions and free binding sites volume fractions along the granule radius at different times are reported in Fig. \ref{f5.4.1.3}. As shown in fig. \ref{f5.4.1.1}, after $20$ days the granule has achieved a radius of about $500 \ \mu m$ and the binding sites of each biofilm component are almost completely consumed. This is ascribed to the combination of different factors: high concentration of metal in the influent wastewater, which is rapidly adsorbed on the granule matrix, granules not completely developed and overall characterized by a small number of binding sites; low fraction of phototrophs throughout the granule. As shown in Fig. \ref{f5.4.1.2}, when the phototrophs and $EPS$ fractions start to be relevant ($T=40 \ d$), the adsorption process is favoured, microorganisms are less inhibited, and the granule radius increases. As a consequence, new free binding sites are formed and immediately occupied. Passing form $T=40 \ d$ to $T=60 \ d$ the granule radius undergoes a further significant increase (Fig. \ref{f5.4.1.1}), and the residual metal concentration is completely adsorbed (see Fig. \ref{f5.4.1.4}) thanks to the high volume fractions of free binding sites. After $80$ days the granule radius, microbial species distribution and volume fractions of free binding sites have achieved the steady-state configuration. Confirming what has been observed in Fig. \ref{f5.4.1.2}, the granule is mainly composed by  $EPS$, phototrophs, inert material, and heterotrophs. The residual binding sites still free indicates the algal-bacterial granules containing in the SBR are perfectly able to remove the metal present in the influent wastewater.  
 
\begin{figure*}    
\centering
\fbox{\includegraphics[width=1\textwidth, keepaspectratio]{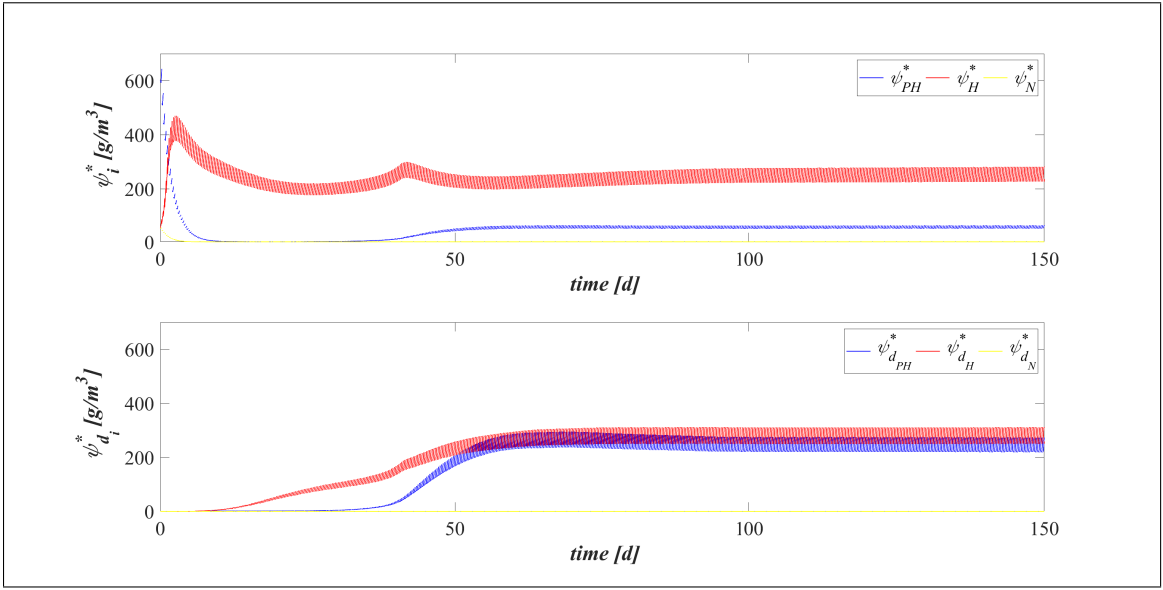}}
\caption{$SET1$ - Evolution of planktonic and detached biomasses concentrations within the reactor over time. Wastewater influent composition: $S^{in}_{IC} = 180 \ g \ m^{-3}$ (inorganic carbon), $S^{in}_{DOC} = 500 \ g \ m^{-3}$ (organic carbon), $S^{in}_{NH_3} = 50 \ g \ m^{-3}$ (ammonia), $S^{in}_{NO_3} = 0$ (nitrate), $S^{in}_{O_2} = 0$ (oxygen), $M^{in}=100 \ g \ m^{-3}$ (metal). Incident light intensity: $I_0= 0.008 \ kmol \ m^{-2} \ d^{-1}$. Duration of the cycle: $\tau =6 \ h$. Time of light exposure: $t_{light} = 50$\% $\tau$.} \label{f5.4.1.6}
\end{figure*}

Fig. \ref{f5.4.1.4} reports the concentration of soluble substrates and metal within the reactor over time. The observation period includes the process start-up until the achievement of the steady-state representative of the working configuration of the reactor. It should be noted that such concentrations have a discontinuous trend, due to the SBR configuration. In the initial phase, the biofilm granules are small, and the consumption and production of soluble substrates are governed by planktonic biomass (see Fig. \ref{f5.4.1.6}). Heterotrophic bacteria and phototrophs have higher growth rates than other microbial species. Thus, $DOC$ (red) and $NH_3$ (cyan) consumption and $IC$ (blue) production can be observed. When photogranules dimension increases, biological processes starts to be governed by sessile species. After $10$ days, the oxygen (magenta) produced by phototrophs in presence of light is not sufficient for heterotrophs and nitrifiers, and no $NO_3$ (yellow) is present in the reactor. Thus, a temporary equilibrium in term of soluble substrates characterizes the system from $10$ to $40$ days. In this time frame, the metal (black) is slowly adsorbed on granules matrix and the metal inhibition effect on metabolic microbial activities reduces over time. As observed in Fig. \ref{f5.4.1.2}, this favours phototrophs growth. Consequently, thanks to $O_2$ production (magenta), heterotrophic bacteria growth is promoted. Moreover, in this phase phototrophs are in turn responsible for more rapid metal adsorption, thanks to their elevate $EPS$ productions and high adsorption capabilities. Metabolic activities of phototrophs and heterotrophs result in the complete $NH_3$ and $DOC$ degradation, $IC$ consumption, and $M$ adsorption after $t=60 \ d$. 

\begin{figure*}    
\centering
\fbox{\includegraphics[width=1\textwidth, keepaspectratio]{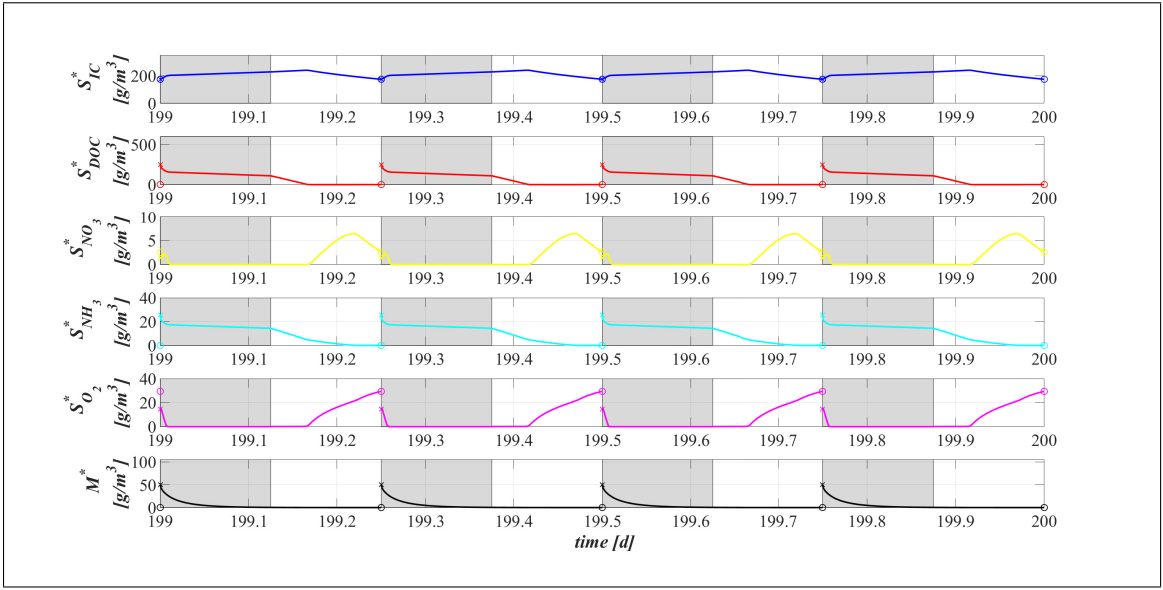}}
\caption{$SET1$ - Evolution of soluble substrates and metal concentrations within the reactor, from $199 \ d$ to $200 \ d$ (four consecutive six-hours treatment cycles). Wastewater influent composition: $S^{in}_{IC} = 180 \ g \ m^{-3}$ (inorganic carbon), $S^{in}_{DOC} = 500 \ g \ m^{-3}$ (organic carbon), $S^{in}_{NH_3} = 50 \ g \ m^{-3}$ (ammonia), $S^{in}_{NO_3} = 0$ (nitrate), $S^{in}_{O_2} = 0$ (oxygen), $M^{in}=100 \ g \ m^{-3}$ (metal). Incident light intensity: $I_0= 0.008 \ kmol \ m^{-2} \ d^{-1}$. Duration of the cycle: $\tau =6 \ h$. Time of light exposure: $t_{light} = 50$\% $\tau$ (grey portions indicate the dark phases, white portions indicate the light phases).} \label{f5.4.1.5}
\end{figure*}  

Once the photogranules have reached a steady-state dimension and microbial stratification, the trend of the substrates and metal concentrations are repeated identically in each cycle. Fig. \ref{f5.4.1.5} shows the evolution of the substrates and metal concentrations over time in the reactor in the period between $199$ and $200$ days. Note that each cycle identically repeats four times in a single day, and, for this reason, it is representative of the operating conditions of the system, while the substrates and metal concentrations at the end of each cycle are representative of the effluent composition. Solid lines represent the trends of $IC$ (blue), $DOC$ (red), $NO_3$ (yellow), $NH_3$ (cyan), $O_2$ (magenta), $M$ (black) concentrations during the cycles. Each cycle is constituted by three hours of dark phase (gray parts of the graphs) and three hours of light phase (white parts of the graphs). While the circle and cross markers represent the concentrations of substrates and metal in the effluent and influent, respectively. At the end of each cycle (at $199.25$, $199.50$, $199.75$, and $200$ days) there is a discontinuity between the inlet and outlet concentration values, due to the procedure of emptying and refilling in the reactor. Due to the absence of light, in the first part of each cycle phototrophs and heterotrophs compete for $O_2$ (produced in the previous cycle), $DOC$ and $NH_3$ producing $IC$. Contextually, a small amount of nitrifiers contributes to the conversion of $O_2$, $IC$ and $NH_3$ into $NO_3$. When oxygen is completely consumed, anoxic heterotrophs grow consuming $DOC$ and $NO_3$. When also the concentration of nitrate reaches zero, the trend of substrate concentrations does not show high variations until the end of the dark period. In light conditions, phototrophs carry out photosynthesis, consuming $NH_3$ and $IC$, and producing large amount of $O_2$ necessary for heterotrophs and nitrifiers. Nevertheless, heterotrophic bacteria are more competitive in the use of $O_2$ in presence of $DOC$. As a result, the $DOC$ concentration reduces and $IC$ concentration increases. When the organic carbon ends, oxygen produced by phototrophs is used by nitrifying bacteria. For this reason, $NH_3$ and $IC$ concentrations decreases, and $NO_3$ concentration increases. When also $NH_3$ is completely consumed, phototrophs grow on $NO_3$. At the end of the cycle, $NH_3$, $DOC$ have been completely removed, and a very low concentrations of $NO_3$ (less than $5 \ g \ m^{-3}$) and a concentration of about $30 \ g \ m^{-3}$ of $O_2$ are observed. Indeed, the biomass of nitrifying bacteria within the granule and their growth rate are very low, therefore the production of $NO_3$ is limited. Regarding the metal adsorption, as already observed in Fig. \ref{f5.4.1.1} (top) at the steady-state the granule is mainly composed by phototrophs, $EPS$ and inert material which are the major responsible for the adsorption process. Thus, during the day/night cycle the metal is completely adsorbed on granule matrix. 
 
The concentration of planktonic and detached biomasses within the reactor over time is shown in Fig. \ref{f5.4.1.6}. As for soluble substrates and metal (Fig. \ref{f5.4.1.4}), the observation period includes the process start-up until the achievement of the steady-state representative of the working configuration of the reactor. Also in this case, such concentrations have a discontinuous trend, due to cyclic behaviour of the SBR. Several factors can affect the suspended biomasses evolution over time. Attachment phenomena contribute to decrease the concentrations of planktonic species, while detachment phenomena promote the growth of detached biomasses. Both types of biomasses grow on soluble substrates and decay. Moreover, their concentration reduces during the emptying phase due to their non perfect settling properties. Lastly, detached biomasses reconvert into the planktonic cells after $48 \ h$ from the detachment. In the initial stage of the process, granules have still small dimension and there is an elevate availability of nutrients. As consequence, the substrates dynamics within the reactor are governed by planktonic biomass. Heterotrophic bacteria have higher growth rate than other microbial species, and their concentration rapidly increases. When the photogranules dimension increases, biological processes are governed by sessile species. Since then, the amount of substrates available for suspended biomass reduces and the concentration of heterotrophs in planktonic form decreases over time. Other species have low growth rates, and their concentrations decrease over time from the beginning of the process, due to the wash-out and attachment process. After $20$ days, photogranules are already formed and the detachment process becomes relevant. This results in the increment of the concentration of  heterotrophic detached biomass. Indeed, as shown in Fig. \ref{f5.4.1.2} the granule is initially composed by large amount of heterotrophs. After $40$ days, sessile phototrophs grow within the granule and, consequently, the concentration of phototrophic detached biomass increases due to the detachment process. The conversion of heterotrophic and phototrophic detached biomasses into planktonic form causes a further increment of planktonic species. Subsequently, due to the shortage of $DOC$ (Fig. \ref{f5.4.1.4}) a slight reduction of planktonic species concentration within the bulk liquid can be observed again. Overall, after $60$ days all suspended species within the reactor achieve a steady state value. 

\subsection{SET2 - Effects of metal concentration on OPGs formation and adsorption processes} \label{n5.4.2}

\begin{figure*}    
\centering
\fbox{\includegraphics[width=1\textwidth, keepaspectratio]{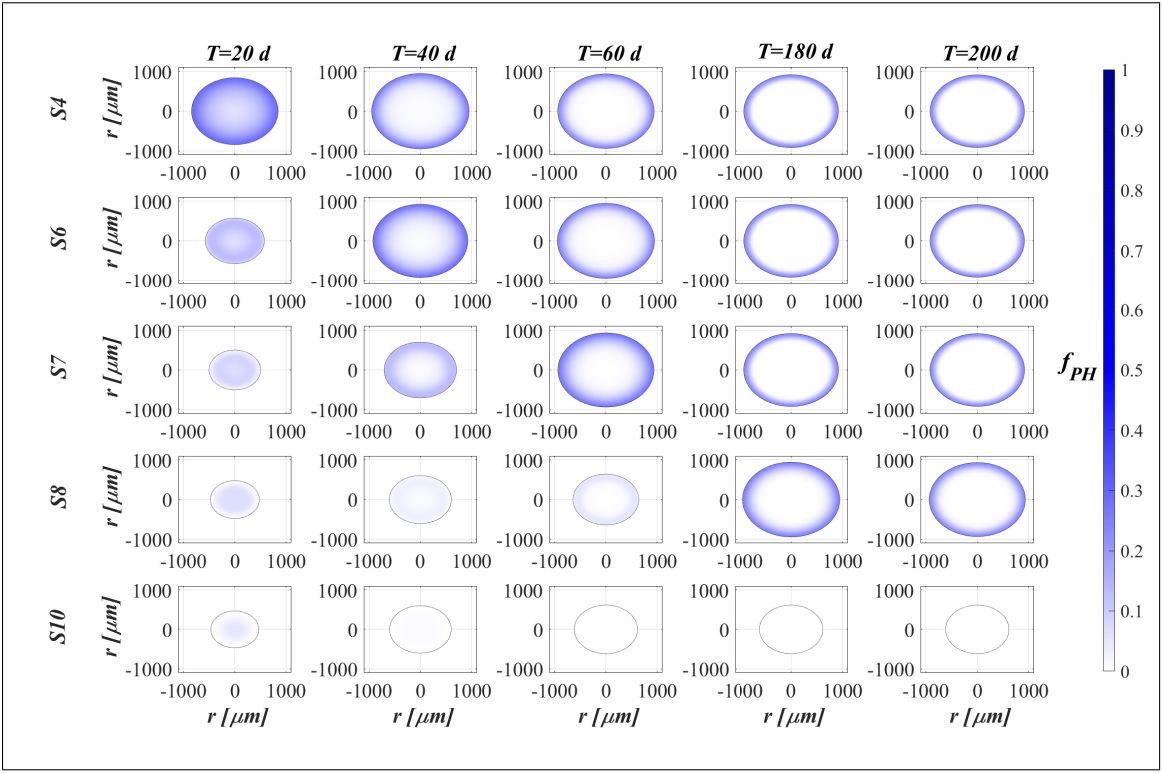}}
\caption{$SET2$ - Phototrophs distribution within the granule (diametrical section) at $T = 20 \ d$, $T = 40 \ d$, $T = 60 \ d$, $T = 180 \ d$, $T = 200 \ d$ for different metal concentrations in the influent $M^{in}$. Wastewater influent composition: $S^{in}_{IC} = 180 \ g \ m^{-3}$ (inorganic carbon), $S^{in}_{DOC} = 500 \ g \ m^{-3}$ (organic carbon), $S^{in}_{NH_3} = 50 \ g \ m^{-3}$ (ammonia), $S^{in}_{NO_3} = 0$ (nitrate), $S^{in}_{O_2} = 0$ (oxygen). $S4: M^{in}=40 \ g \ m^{-3}$, $S6: M^{in}=80 \ g \ m^{-3}$, $S7: M^{in}=100 \ g \ m^{-3}$, $S8: M^{in}=120 \ g \ m^{-3}$, $S10: M^{in}=160 \ g \ m^{-3}$. Incident light intensity: $I_0= 0.008 \ kmol \ m^{-2} \ d^{-1}$. Duration of the cycle: $\tau =6 \ h$. Time of light exposure: $t_{light} = 50$\% $\tau$.} \label{f5.4.2.1}
\end{figure*}

\begin{figure*}    
\centering
\fbox{\includegraphics[width=1\textwidth, keepaspectratio]{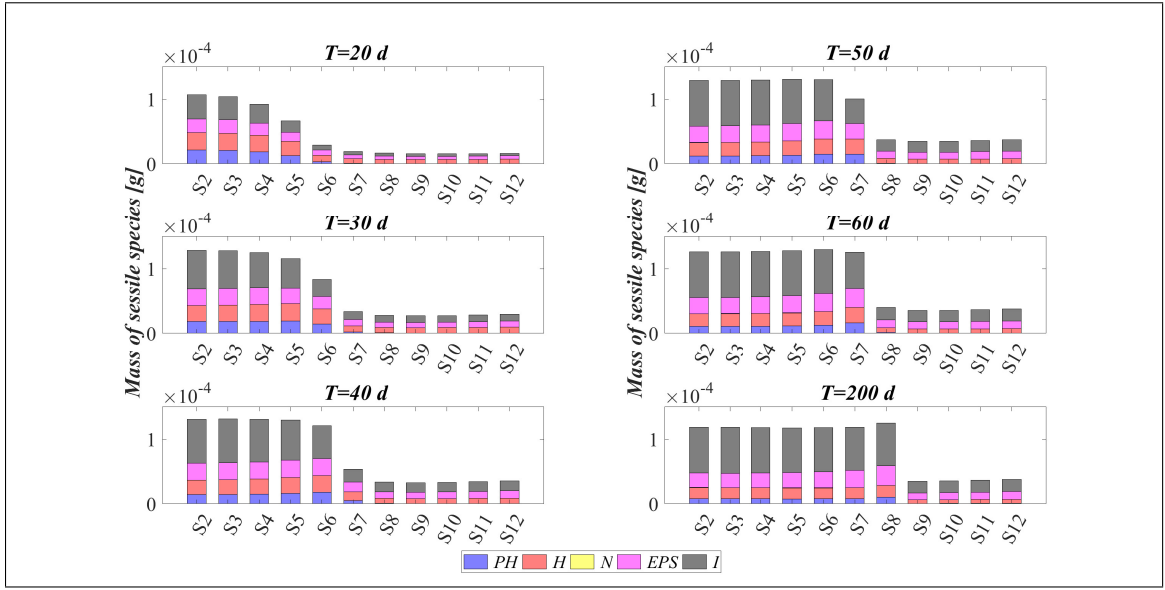}}
\caption{$SET2$ - Mass of microbial species within the granule at $T = 20 \ d$, $T = 30 \ d$, $T = 40 \ d$, $T = 50 \ d$, $T = 60 \ d$, $T = 200 \ d$ for different metal concentrations in the influent $M^{in}$. Wastewater influent composition: $S^{in}_{IC} = 180 \ g \ m^{-3}$ (inorganic carbon), $S^{in}_{DOC} = 500 \ g \ m^{-3}$ (organic carbon), $S^{in}_{NH_3} = 50 \ g \ m^{-3}$ (ammonia), $S^{in}_{NO_3} = 0$ (nitrate), $S^{in}_{O_2} = 0$ (oxygen). $S2: M^{in}=0$, $S3: M^{in}=20 \ g \ m^{-3}$, $S4: M^{in}=40 \ g \ m^{-3}$, $S5: M^{in}=60 \ g \ m^{-3}$, $S6: M^{in}=80 \ g \ m^{-3}$, $S7: M^{in}=100 \ g \ m^{-3}$, $S8: M^{in}=120 \ g \ m^{-3}$, $S9: M^{in}=140 \ g \ m^{-3}$, $S10: M^{in}=160 \ g \ m^{-3}$, $S11: M^{in}=180 \ g \ m^{-3}$, $S12: M^{in}=200 \ g \ m^{-3}$. Incident light intensity: $I_0= 0.008 \ kmol \ m^{-2} \ d^{-1}$. Duration of the cycle: $\tau =6 \ h$. Time of light exposure: $t_{light} = 50$\% $\tau$.} \label{f5.4.2.2}
\end{figure*}

\begin{figure*}    
\centering
\fbox{\includegraphics[width=1\textwidth, keepaspectratio]{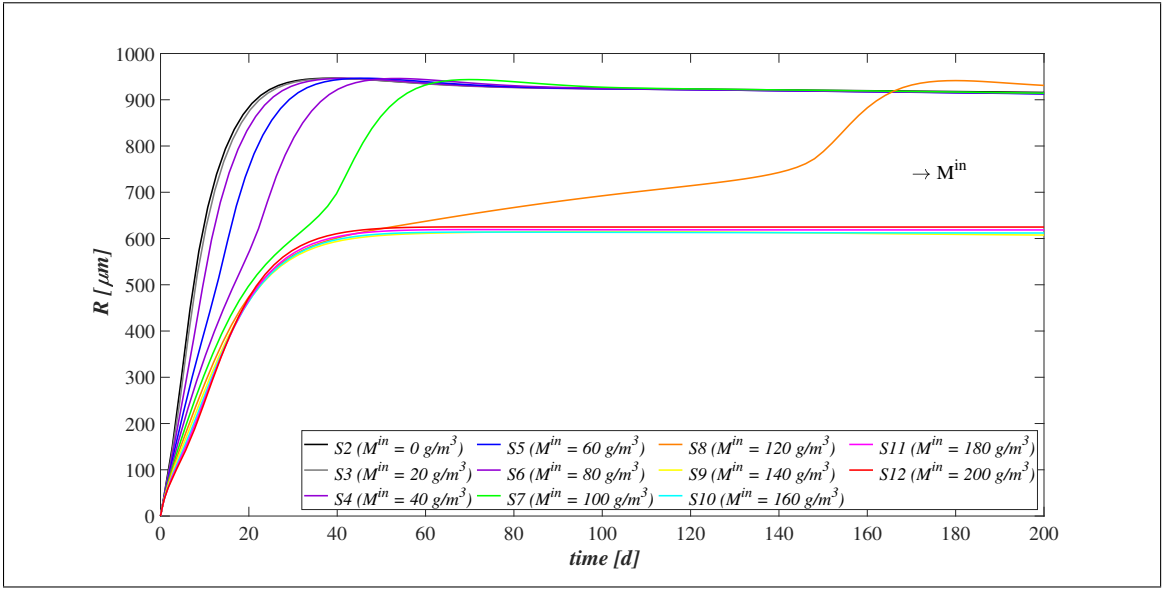}}
\caption{$SET2$ - Biofilm radius evolution over time for different metal concentrations in the influent $M^{in}$. Wastewater influent composition: $S^{in}_{IC} = 180 \ g \ m^{-3}$ (inorganic carbon), $S^{in}_{DOC} = 500 \ g \ m^{-3}$ (organic carbon), $S^{in}_{NH_3} = 50 \ g \ m^{-3}$ (ammonia), $S^{in}_{NO_3} = 0$ (nitrate), $S^{in}_{O_2} = 0$ (oxygen). $S2: M^{in}=0$, $S3: M^{in}=20 \ g \ m^{-3}$, $S4: M^{in}=40 \ g \ m^{-3}$, $S5: M^{in}=60 \ g \ m^{-3}$, $S6: M^{in}=80 \ g \ m^{-3}$, $S7: M^{in}=100 \ g \ m^{-3}$, $S8: M^{in}=120 \ g \ m^{-3}$, $S9: M^{in}=140 \ g \ m^{-3}$, $S10: M^{in}=160 \ g \ m^{-3}$, $S11: M^{in}=180 \ g \ m^{-3}$, $S12: M^{in}=200 \ g \ m^{-3}$. Incident light intensity: $I_0= 0.008 \ kmol \ m^{-2} \ d^{-1}$. Duration of the cycle: $\tau =6 \ h$. Time of light exposure: $t_{light} = 50$\% $\tau$.} \label{f5.4.2.3}
\end{figure*}

\begin{figure*}    
\centering
\fbox{\includegraphics[width=1\textwidth, keepaspectratio]{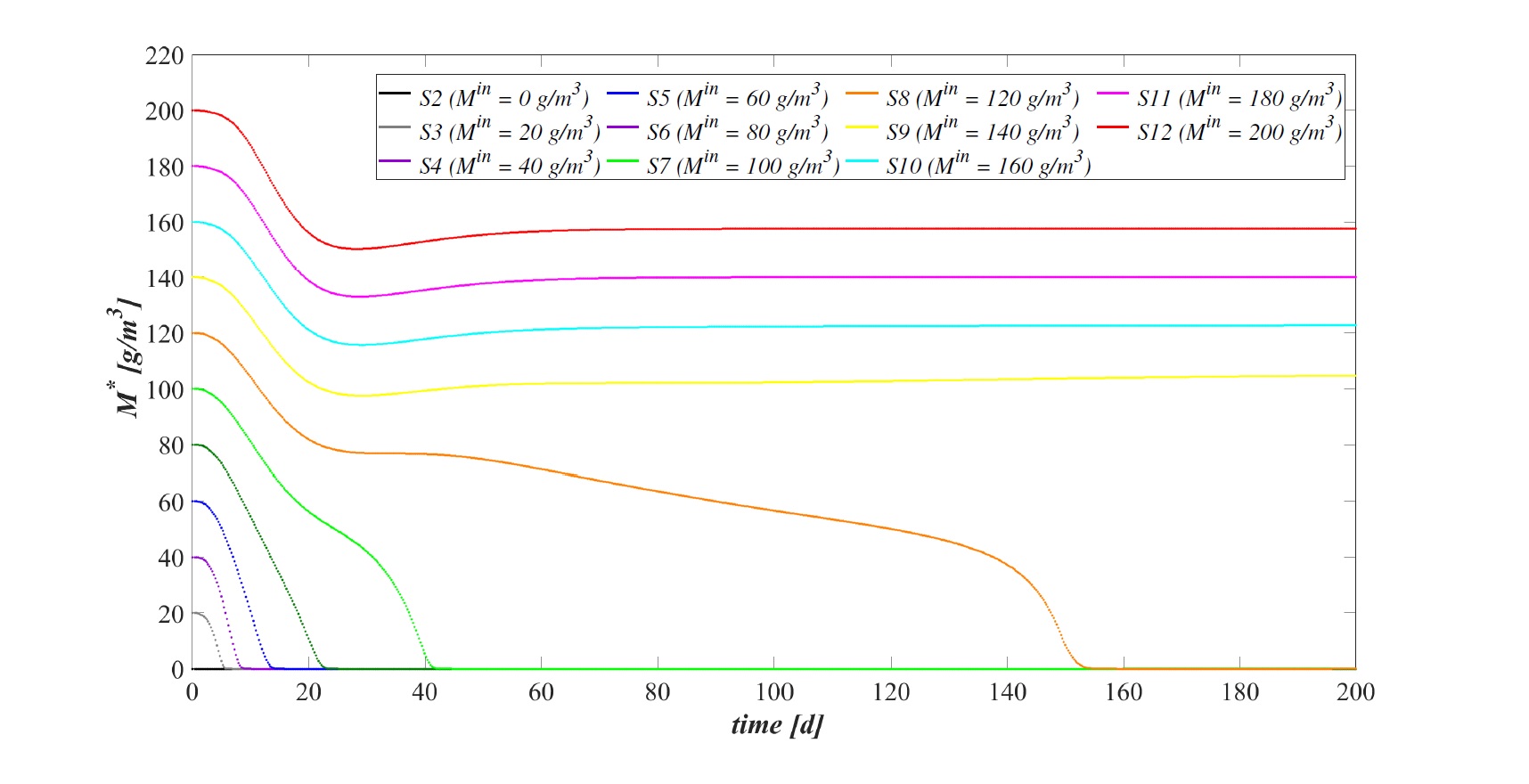}}
\caption{$SET2$ - Metal concentration evolution within the reactor for different metal concentrations in the influent $M^{in}$. Wastewater influent composition: $S^{in}_{IC} = 180 \ g \ m^{-3}$ (inorganic carbon), $S^{in}_{DOC} = 500 \ g \ m^{-3}$ (organic carbon), $S^{in}_{NH_3} = 50 \ g \ m^{-3}$ (ammonia), $S^{in}_{NO_3} = 0$ (nitrate), $S^{in}_{O_2} = 0$ (oxygen). $S2: M^{in}=0$, $S3: M^{in}=20 \ g \ m^{-3}$, $S4: M^{in}=40 \ g \ m^{-3}$, $S5: M^{in}=60 \ g \ m^{-3}$, $S6: M^{in}=80 \ g \ m^{-3}$, $S7: M^{in}=100 \ g \ m^{-3}$, $S8: M^{in}=120 \ g \ m^{-3}$, $S9: M^{in}=140 \ g \ m^{-3}$, $S10: M^{in}=160 \ g \ m^{-3}$, $S11: M^{in}=180 \ g \ m^{-3}$, $S12: M^{in}=200 \ g \ m^{-3}$. Incident light intensity: $I_0= 0.008 \ kmol \ m^{-2} \ d^{-1}$. Duration of the cycle: $\tau =6 \ h$. Time of light exposure: $t_{light} = 50$\% $\tau$.} \label{f5.4.2.4}
\end{figure*}

\begin{figure*}    
\centering
\fbox{\includegraphics[width=1\textwidth, keepaspectratio]{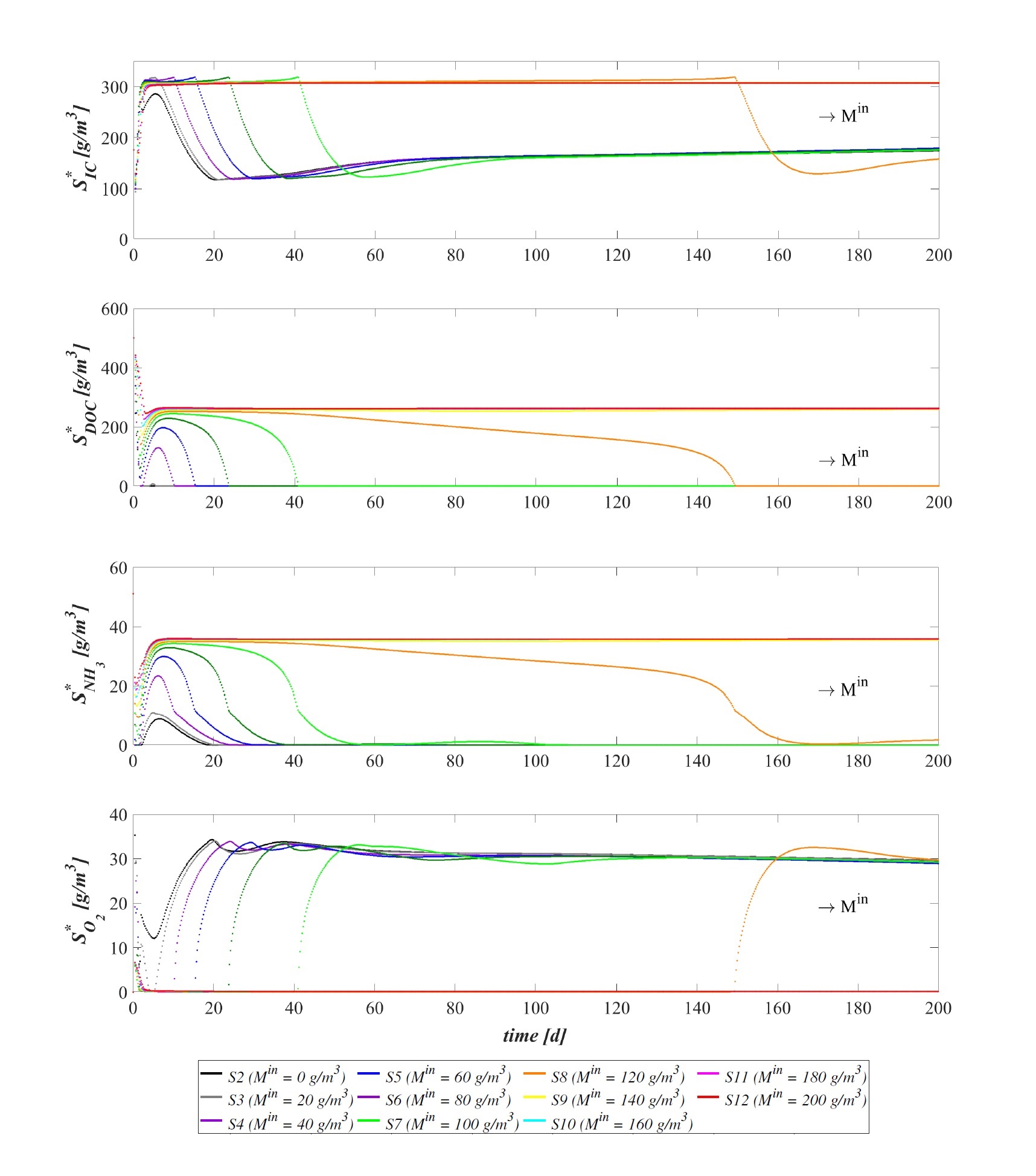}}
\caption{$SET2$ - Evolution of soluble substrates concentration within the reactor for different metal concentrations in the influent $M^{in}$. Wastewater influent composition: $S^{in}_{IC} = 180 \ g \ m^{-3}$ (inorganic carbon), $S^{in}_{DOC} = 500 \ g \ m^{-3}$ (organic carbon), $S^{in}_{NH_3} = 50 \ g \ m^{-3}$ (ammonia), $S^{in}_{NO_3} = 0$ (nitrate), $S^{in}_{O_2} = 0 \ g \ m^{-3}$ (oxygen). $S2: M^{in}=0$, $S3: M^{in}=20 \ g \ m^{-3}$, $S4: M^{in}=40 \ g \ m^{-3}$, $S5: M^{in}=60 \ g \ m^{-3}$, $S6: M^{in}=80 \ g \ m^{-3}$, $S7: M^{in}=100 \ g \ m^{-3}$, $S8: M^{in}=120 \ g \ m^{-3}$, $S9: M^{in}=140 \ g \ m^{-3}$, $S10: M^{in}=160 \ g \ m^{-3}$, $S11: M^{in}=180 \ g \ m^{-3}$, $S12: M^{in}=200 \ g \ m^{-3}$.  Incident light intensity: $I_0= 0.008 \ kmol \ m^{-2} \ d^{-1}$. Duration of the cycle: $\tau =6 \ h$. Time of light exposure: $t_{light} = 50$\% $\tau$.} \label{f5.4.2.5}
\end{figure*}

\begin{figure*}    
\centering
\fbox{\includegraphics[width=1\textwidth, keepaspectratio]{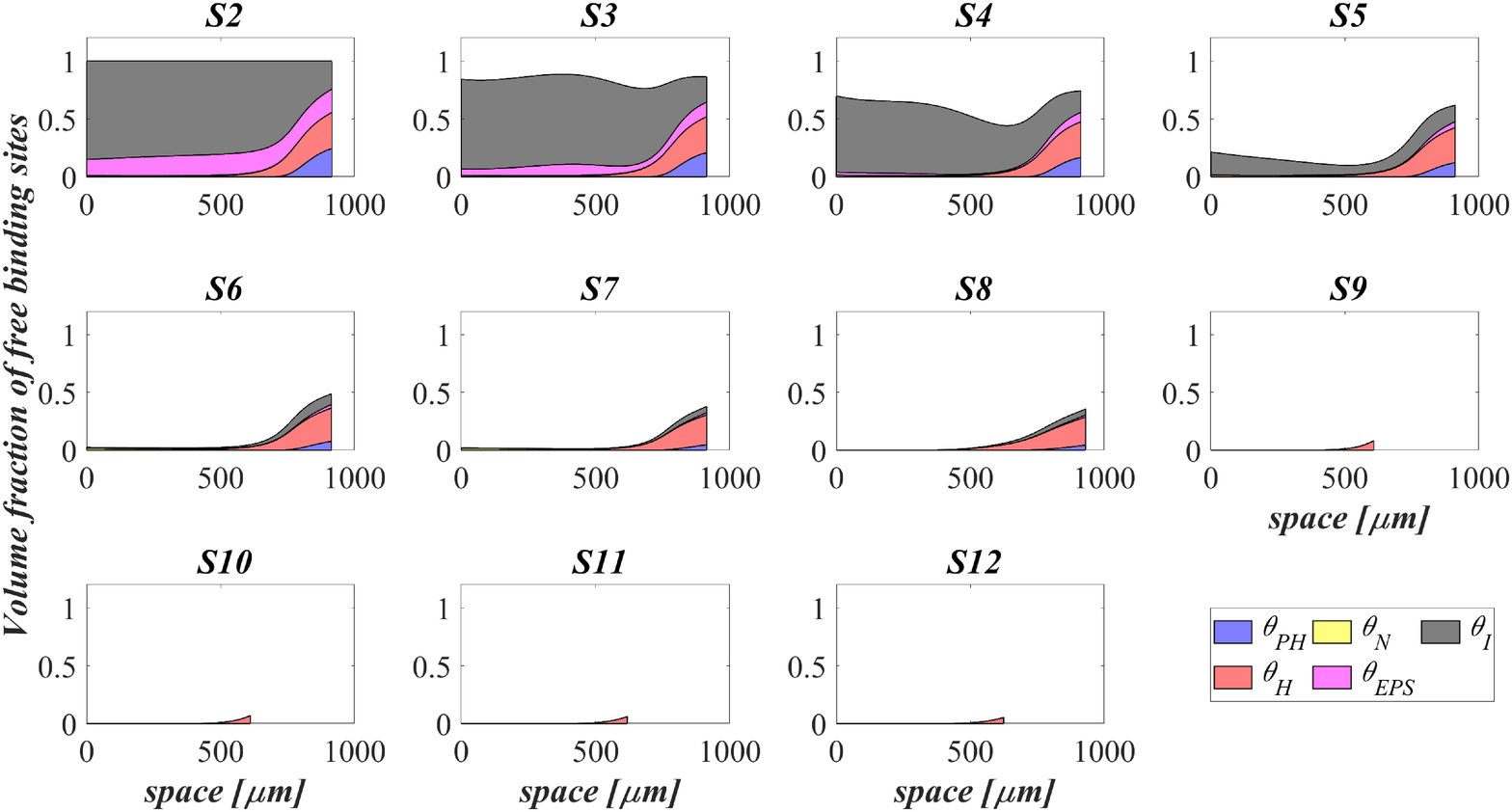}}
\caption{$SET2$ - Distribution of the residual free binding site across the radius of the granule for different metal concentrations in the influent $M^{in}$ at $T=200 \ d$. Wastewater influent composition: $S^{in}_{IC} = 180 \ g \ m^{-3}$ (inorganic carbon), $S^{in}_{DOC} = 500 \ g \ m^{-3}$ (organic carbon), $S^{in}_{NH_3} = 50 \ g \ m^{-3}$ (ammonia), $S^{in}_{NO_3} = 0$ (nitrate), $S^{in}_{O_2} = 0$ (oxygen). $S2: M^{in}=0$, $S3: M^{in}=20 \ g \ m^{-3}$, $S4: M^{in}=40 \ g \ m^{-3}$, $S5: M^{in}=60 \ g \ m^{-3}$, $S6: M^{in}=80 \ g \ m^{-3}$, $S7: M^{in}=100 \ g \ m^{-3}$, $S8: M^{in}=120 \ g \ m^{-3}$, $S9: M^{in}=140 \ g \ m^{-3}$, $S10: M^{in}=160 \ g \ m^{-3}$, $S11: M^{in}=180 \ g \ m^{-3}$, $S12: M^{in}=200 \ g \ m^{-3}$. Incident light intensity: $I_0= 0.008 \ kmol \ m^{-2} \ d^{-1}$. Duration of the cycle: $\tau =6 \ h$. Time of light exposure: $t_{light} = 50$\% $\tau$.} \label{f5.4.2.6}
\end{figure*}

Metals in wastewater may increase the sessile production of $EPS$, and, at the same time, may be the cause of stress conditions responsible for the death of microbial cells. More studies are necessary to identify a concentration range that allows microorganisms to grow and secrete $EPS$ maximizing the removal efficiency of metals from wastewater. In this numerical study $SET2$, the efficiency of metal adsorption on the matrix of biofilm granules and the inhibiting effect on OPGs formation are investigated by considering different concentrations of metal. For this purpose, eleven simulations ($S2$ - $S12$) have been carried out by setting the concentration of metal in the influent $M^{in}$ equal to $0$, $20$, $40$, $60$, $80$, $100$, $120$, $140$, $160$, $180$, $200 \ g \ m^{-3}$. The concentration of soluble substrates in the influent wastewater $S^{in}_j$ and initial concentration of planktonic biomasses within the reactor $\psi^*_{i,0}$ set for this numerical study are the same as in $SET1$. Numerical results are summarized in Figs. \ref{f5.4.2.1}-\ref{f5.4.2.6}. 

The concentration of metal in the influent affects the adsorption process as well as the microbial species stratification of OPGs. The distribution of the phototrophic sessile biomass at different times is reported in Fig. \ref{f5.4.2.1} for the following simulations: $S4$ ($M^{in}=40 \ g \ m^{-3}$), $S6$ ($M^{in}=80 \ g \ m^{-3}$), $S7$ ($M^{in}=100 \ g \ m^{-3}$), $S8$ ($M^{in}=120 \ g \ m^{-3}$), $S10$ ($M^{in}=160 \ g \ m^{-3}$). When the metal concentration present in the bioreactor is very low ($S4$), phototrophs are less inhibited and grow faster within the granule. Conversely, a higher concentration of free metal results in a higher inhibition effect and leads to a slower growth of phototrophs. It means that for wastewater richer in metal the growth of the phototrophic species and the adsorption process occur in a longer time. Thus, the maximum fraction of phototrophs is observed later going from $S4$ to $S8$. However, after long times the phototrophs distribution is no longer affected by $M^{in}$ and all simulations achieve the same steady-state configuration after $200$ days, except for $M^{in}=160 \ g \ m^{-3}$. Indeed, in this case ($S10$) the metal concentration is too high, and the biomasses growth and the granule formation are strongly inhibited by the presence of free metal. As a result, the photogranule does not completely develop and the absence of phototrophs is observed throughout the granule at $200$ days.

This is visible also in Fig. \ref{f5.4.2.2}, where the mass of sessile microbial species within the granule is shown at different times. Relevant differences concern the initial phase of the process when the total sessile mass is higher for low metal concentrations. However, after long times (when the adsorption process is completed) the sessile mass of the individual microbial species within the granule is no longer affected by the presence of metal for $M^{in}$ lower than $140 \ g \ m^{-3}$.  

The evolution of the granule radius $R(t)$ over time is shown in Fig. \ref{f5.4.2.3}. $M^{in}$ affects the granule evolution in the initial stage of the process. For low concentrations of metal, the granule radius increases earlier, small inhibiting effects are observed during the granulation process, and the adsorption process is faster completed. When $M^{in}$ is higher or equal to $100 \ g \ m^{-3}$, the granule partially grows, and a further radius increment associated to phototrophs growth can be observed later. Note that passing from $S6$ to $S8$ the phototrophs growth and the subsequent further increment of the granule radius are increasingly slowed down by the presence of free metal, as observed in Fig. \ref{f5.4.2.1}. Instead, when $M^{in}$ is equal to $140$, $160$, $180$ and $200 \ g \ m^{-3}$ the phototrophic biomass growth is totally inhibited. Hence, the absence of phototrophs observed in Fig. \ref{f5.4.2.1} and Fig. \ref{f5.4.2.2} is confirmed by the incomplete development of the granule.

The metal and substrates concentrations in the effluent after each SBR cycle are displayed in Fig. \ref{f5.4.2.4} and \ref{f5.4.2.5}. Each point represents the concentrations of metal and substrates in the effluent at the end of each cycle. Fig. \ref{f5.4.2.4} shows that for low metal concentration in the influent wastewater, the adsorption process is rapid and almost linear. When $M^{in}$ is equal to $100$ and $120 \ g \ m^{-3}$, the removal process slows down at $T=20 \ d$. The phototrophic biomass is characterized by lower growth rate than heterotrophs, and the inhibiting effect related to the presence of free metal further limits their growth process. Later, after $30$ days, a considerable growth of phototrophs allows to successfully complete the adsorption process. Specifically, higher is the metal concentration in the influent and slower is the adsorption process. This is related to two aspects: the amount of free metal to adsorb is larger, and the phototrophs growth is slower due to the stronger metal inhibition effect. As shown in Fig. \ref{f5.4.2.2}, for $M^{in}$ higher or equal to $140 \ g \ m^{-3}$ the granule is not completely developed, and a small amount of $EPS$ and a not visible fraction of phototrophs are present in the granule. Consequently, the adsorption process is not yet completed after $200$ days. 

The effluent concentration of soluble substrates is reported in Fig. \ref{f5.4.2.5}. The initial phase of the process, in which $DOC$ and $NH_3$ are consumed and $IC$ is produced, is faster when the metal concentration in the influent is lower. Indeed, the inhibiting effect delays the granulation process and, as consequence, the substrates consumption/production. The subsequent phase is governed by phototrophs, which promote the metabolic activities of the heterotrophic bacteria producing $O_2$. When  $M^{in}$ is higher, phototrophs growth is slower, and oxygen is less rapidly produced. As a result, time necessary to reach the complete degradation of $DOC$ and $NH_3$ and the maximum $IC$ reduction increases. Specifically, the same effluent composition both in terms of metal (Fig.  \ref{f5.4.2.4}) and substrates (Fig. \ref{f5.4.2.5}) is achieved at the steady-state for $M^{in}$ lower than $140 \ g \ m^{-3}$. While, in the other cases the degradation of nutrients and metal adsorption only partially occur.  
 
The steady-state configuration of the residual free binding sites within the biofilm granule is reported in Fig. \ref{f5.4.2.6}. Obviously, when no metal is present in the influent wastewater ($S2$), no adsorption site is occupied during the granulation process and the sum of all volume fractions returns 1 at each location and time. The numerical results show that, for values of $M^{in}$ between $20$ and $120 \ g \ m^{-3}$, the adsorption process requires a growing number of binding sites. Consequently, a decreasing residual amount of adsorption sites can be observed, although the granule achieves the same steady-state dimension. Lastly, in the cases in which the metal concentration is too high, the adsorption process is not completed (from $S9$ to $S12$), the granule does not completely develop due to the stronger inhibiting effect, and the binding sites are almost occupied.

\subsection{SET3 - Effects of metal adsorption capabilities on OPGs formation and adsorption processes} \label{n5.4.3}

Experimental works show that heat or acid pretreatments enhance the metal adsorption potential of algal-bacterial biomass \cite{mehta2005use}. Metal affinity to the biomass could be manipulated by pretreating the biomass with alkalies, acids, detergents, and heat, which may increase the amount of adsorbed metal and reduce the time necessary for the adsorption process \cite{ahalya2003biosorption}. Indeed, a physical/chemical pretreatment affects the permeability and surface charge of the biomass and makes the adsorption sites more accessible for metal biosorption \cite{abbas2014biosorption}. In this context, a numerical study ($SET3$) is performed to investigate the pretreatment effect on the evolution of biofilm granules and metal removal. Five simulations ($S13$-$S17$) have been carried out using different values of binding sites densities $\rho_{\theta}$.
 
 \begin{figure*}    
\centering
\fbox{\includegraphics[width=1\textwidth, keepaspectratio]{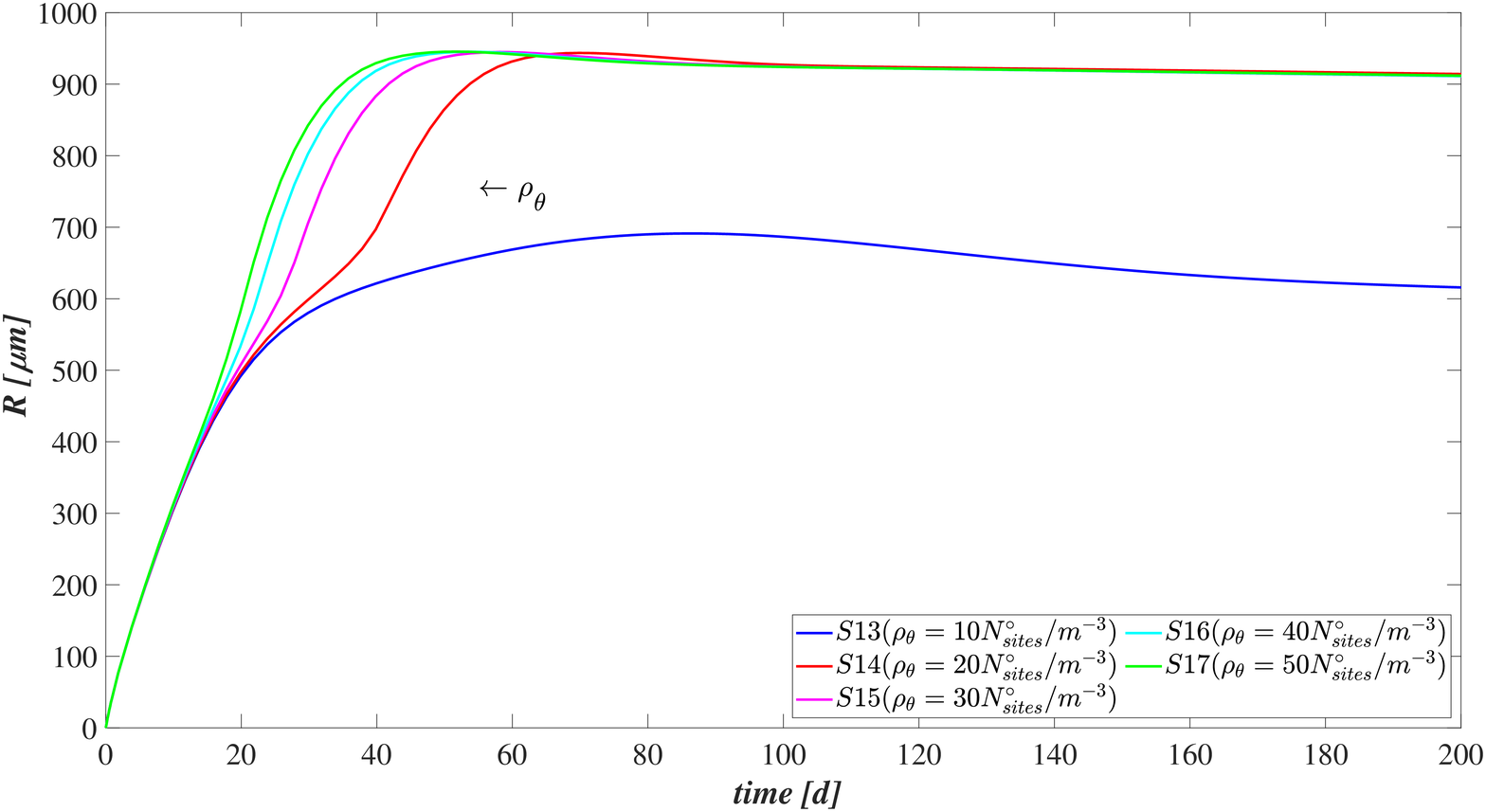}}
\caption{$SET3$ - Biofilm radius evolution over time for different densities of binding sites $\rho_{\theta}$. Wastewater influent composition: $S^{in}_{IC} = 180 \ g \ m^{-3}$ (inorganic carbon), $S^{in}_{DOC} = 500 \ g \ m^{-3}$ (organic carbon), $S^{in}_{NH_3} = 50 \ g \ m^{-3}$ (ammonia), $S^{in}_{NO_3} = 0$ (nitrate), $S^{in}_{O_2} = 0$ (oxygen). $S13: \rho_{\theta}=10 \ N^{\circ}_{sites} \ m^{-3}$, $S14: \rho_{\theta}=20 \ N^{\circ}_{sites} \ m^{-3}$, $S15: \rho_{\theta}=30 \ N^{\circ}_{sites} \ m^{-3}$, $S16: \rho_{\theta}=40 \ N^{\circ}_{sites} \ m^{-3}$, $S17: \rho_{\theta}=50 \ N^{\circ}_{sites} \ m^{-3}$. Incident light intensity: $I_0= 0.008 \ kmol \ m^{-2} \ d^{-1}$. Duration of the cycle: $\tau =6 \ h$. Time of light exposure: $t_{light} = 50$\% $\tau$.} \label{f5.4.3.1}
\end{figure*}  

\begin{figure*}    
\centering
\fbox{\includegraphics[width=1\textwidth, keepaspectratio]{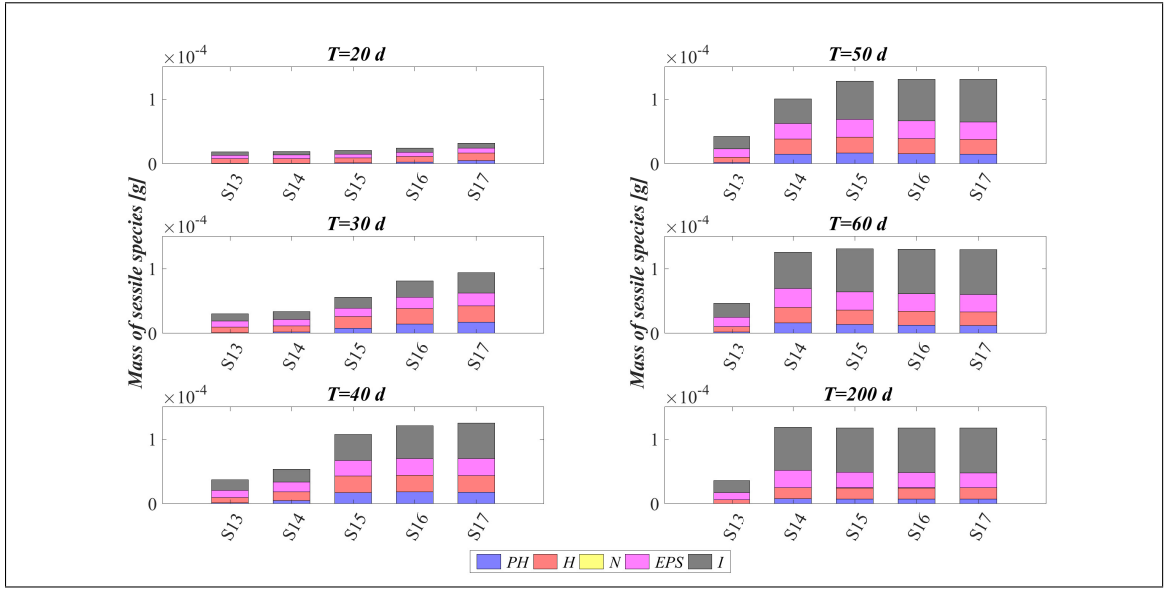}}
\caption{$SET3$ - Mass of microbial species within the granule at $T = 20 \ d$, $T = 30 \ d$, $T = 40 \ d$, $T = 50 \ d$, $T = 60 \ d$, $T = 200 \ d$ for different densities of binding sites $\rho_{\theta}$. Wastewater influent composition: $S^{in}_{IC} = 180 \ g \ m^{-3}$ (inorganic carbon), $S^{in}_{DOC} = 500 \ g \ m^{-3}$ (organic carbon), $S^{in}_{NH_3} = 50 \ g \ m^{-3}$ (ammonia), $S^{in}_{NO_3} = 0$ (nitrate), $S^{in}_{O_2} = 0$ (oxygen). $S13: \rho_{\theta}=10 \ N^{\circ}_{sites} \ m^{-3}$, $S14: \rho_{\theta}=20 \ N^{\circ}_{sites} \ m^{-3}$, $S15: \rho_{\theta}=30 \ N^{\circ}_{sites} \ m^{-3}$, $S16: \rho_{\theta}=40 \ N^{\circ}_{sites} \ m^{-3}$, $S17: \rho_{\theta}=50 \ N^{\circ}_{sites} \ m^{-3}$.  Incident light intensity: $I_0= 0.008 \ kmol \ m^{-2} \ d^{-1}$. Duration of the cycle: $\tau =6 \ h$. Time of light exposure: $t_{light} = 50$\% $\tau$.} \label{f5.4.3.2}
\end{figure*}

\begin{figure*}    
\centering
\fbox{\includegraphics[width=1\textwidth, keepaspectratio]{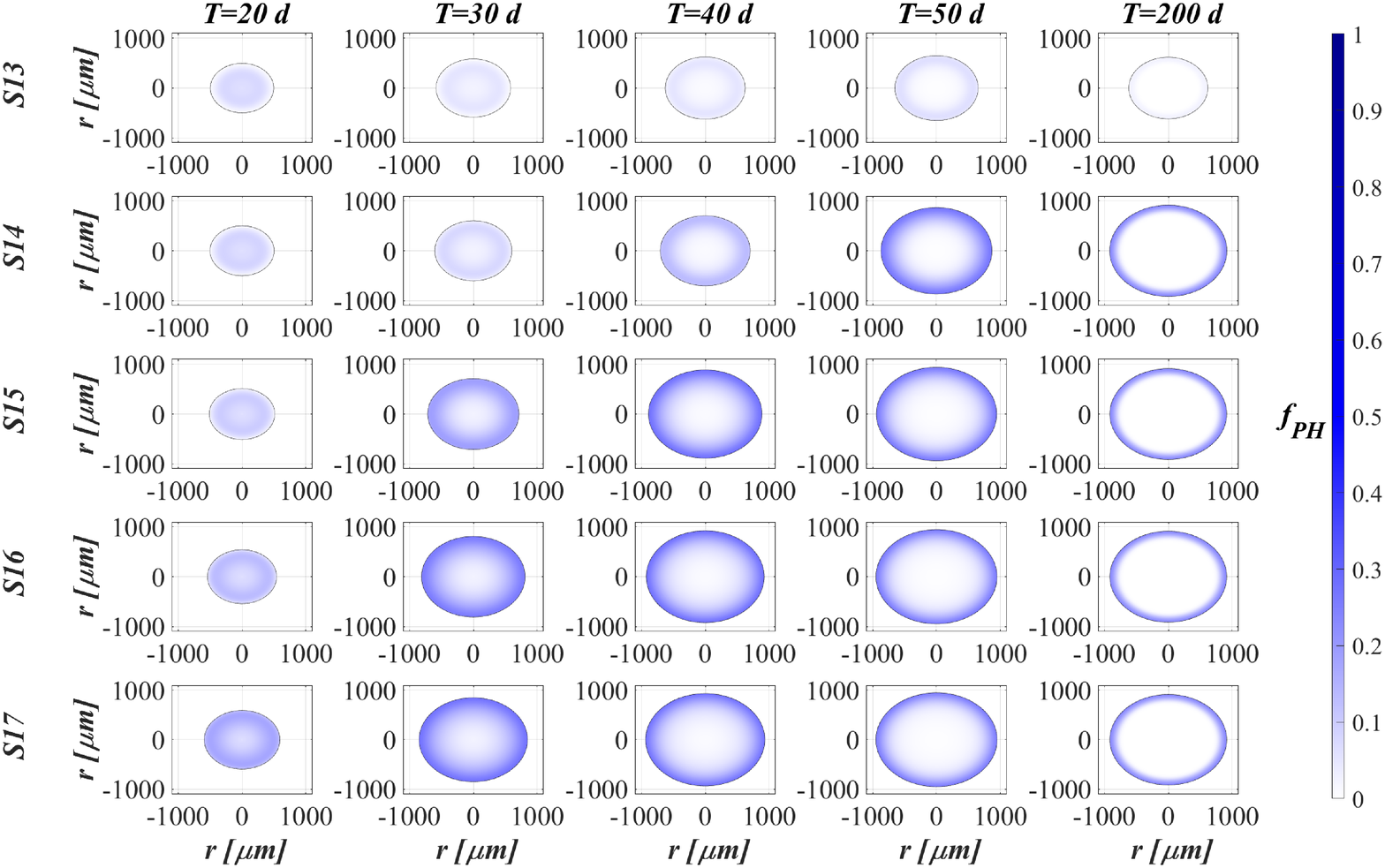}}
\caption{$SET3$ - Phototrophs distribution within the granule (diametrical section) at $T = 20 \ d$, $T = 30 \ d$, $T = 40 \ d$, $T = 50 \ d$, $T = 200 \ d$ for different densities of binding sites $\rho_{\theta}$. Wastewater influent composition: $S^{in}_{IC} = 180 \ g \ m^{-3}$ (inorganic carbon), $S^{in}_{DOC} = 500 \ g \ m^{-3}$ (organic carbon), $S^{in}_{NH_3} = 50 \ g \ m^{-3}$ (ammonia), $S^{in}_{NO_3} = 0$ (nitrate), $S^{in}_{O_2} = 0$ (oxygen). $S13: \rho_{\theta}=10 \ N^{\circ}_{sites} \ m^{-3}$, $S14: \rho_{\theta}=20 \ N^{\circ}_{sites} \ m^{-3}$, $S15: \rho_{\theta}=30 \ N^{\circ}_{sites} \ m^{-3}$, $S16: \rho_{\theta}=40 \ N^{\circ}_{sites} \ m^{-3}$, $S17: \rho_{\theta}=50 \ N^{\circ}_{sites} \ m^{-3}$. Incident light intensity: $I_0= 0.008 \ kmol \ m^{-2} \ d^{-1}$. Duration of the cycle: $\tau =6 \ h$. Time of light exposure: $t_{light} = 50$\% $\tau$.} \label{f5.4.3.3}
\end{figure*}

\begin{figure*}    
\centering
\fbox{\includegraphics[width=1\textwidth, keepaspectratio]{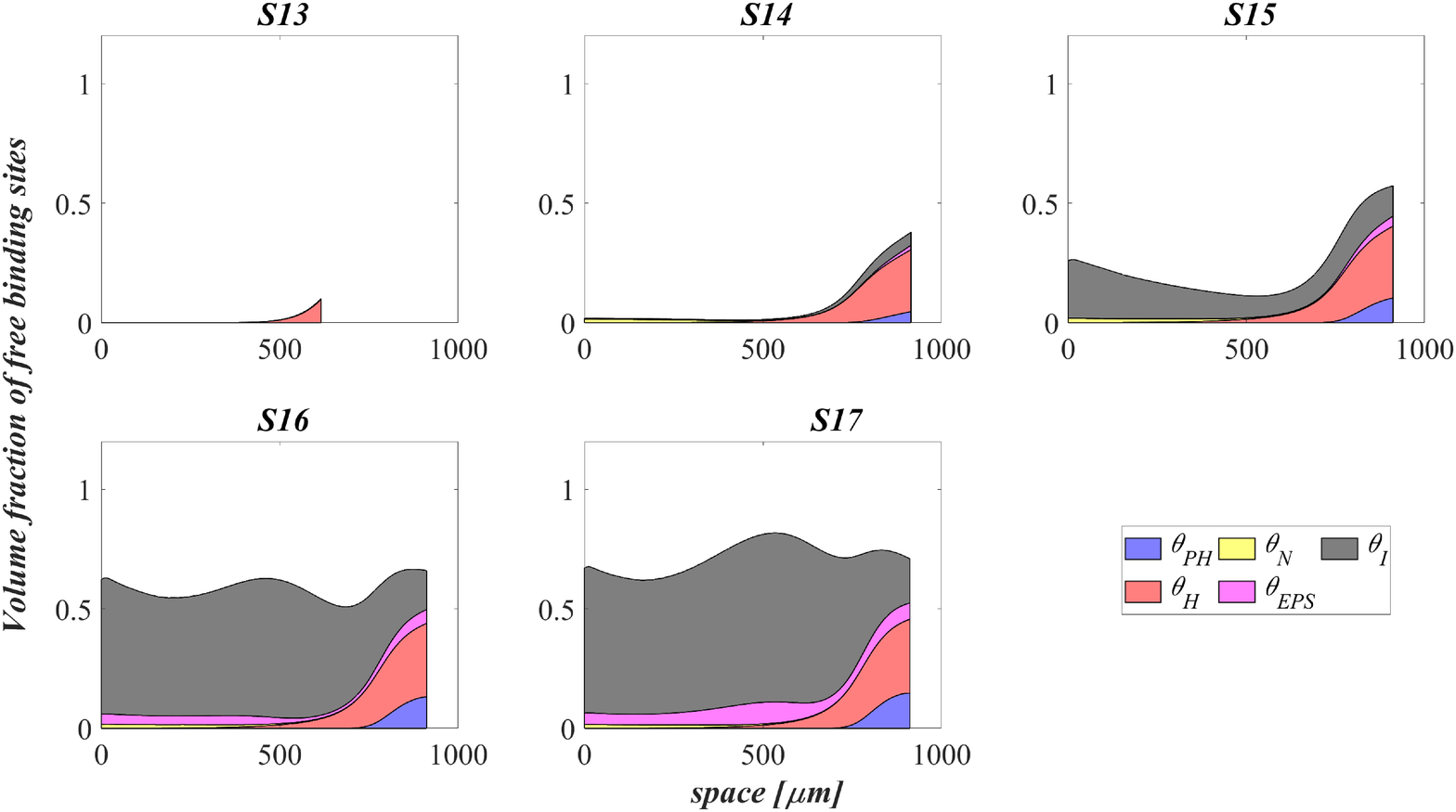}}
\caption{$SET3$ - Distribution of the residual free binding site across the radius of the granule for different densities of binding sites $\rho_{\theta}$ at $T=200 \ d$. Wastewater influent composition: $S^{in}_{IC} = 180 \ g \ m^{-3}$ (inorganic carbon), $S^{in}_{DOC} = 500 \ g \ m^{-3}$ (organic carbon), $S^{in}_{NH_3} = 50 \ g \ m^{-3}$ (ammonia), $S^{in}_{NO_3} = 0$ (nitrate), $S^{in}_{O_2} = 0$ (oxygen). $S13: \rho_{\theta}=10 \ N^{\circ}_{sites} \ m^{-3}$, $S14: \rho_{\theta}=20 \ N^{\circ}_{sites} \ m^{-3}$, $S15: \rho_{\theta}=30 \ N^{\circ}_{sites} \ m^{-3}$, $S16: \rho_{\theta}=40 \ N^{\circ}_{sites} \ m^{-3}$, $S17: \rho_{\theta}=50 \ N^{\circ}_{sites} \ m^{-3}$. Incident light intensity: $I_0= 0.008 \ kmol \ m^{-2} \ d^{-1}$. Duration of the cycle: $\tau =6 \ h$. Time of light exposure: $t_{light} = 50$\% $\tau$.} \label{f5.4.3.6}
\end{figure*}

\begin{figure*}    
\centering
\fbox{\includegraphics[width=1\textwidth, keepaspectratio]{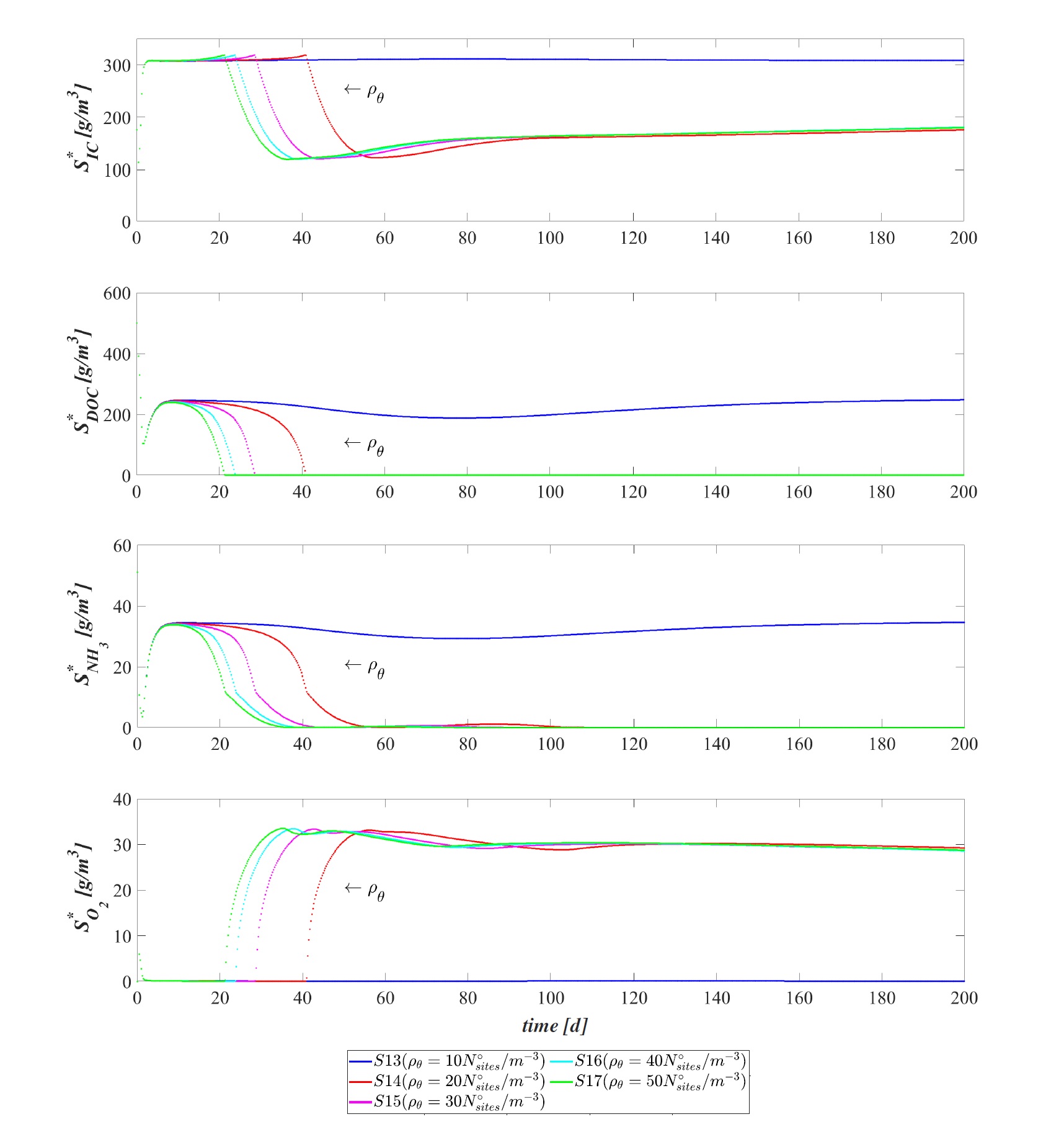}}
\caption{$SET3$ - Soluble substrates concentration evolution within the reactor for different densities of binding sites $\rho_{\theta}$. Wastewater influent composition: $S^{in}_{IC} = 180 \ g \ m^{-3}$ (inorganic carbon), $S^{in}_{DOC} = 500 \ g \ m^{-3}$ (organic carbon), $S^{in}_{NH_3} = 50 \ g \ m^{-3}$ (ammonia), $S^{in}_{NO_3} = 0$ (nitrate), $S^{in}_{O_2} = 0$ (oxygen). $S13: \rho_{\theta}=10 \ N^{\circ}_{sites} \ m^{-3}$, $S14: \rho_{\theta}=20 \ N^{\circ}_{sites} \ m^{-3}$, $S15: \rho_{\theta}=30 \ N^{\circ}_{sites} \ m^{-3}$, $S16: \rho_{\theta}=40 \ N^{\circ}_{sites} \ m^{-3}$, $S17: \rho_{\theta}=50 \ N^{\circ}_{sites} \ m^{-3}$. Incident light intensity: $I_0= 0.008 \ kmol \ m^{-2} \ d^{-1}$. Duration of the cycle: $\tau =6 \ h$. Time of light exposure: $t_{light} = 50$\% $\tau$.} \label{f5.4.3.5}
\end{figure*}

\begin{figure*}    
\centering
\fbox{\includegraphics[width=1\textwidth, keepaspectratio]{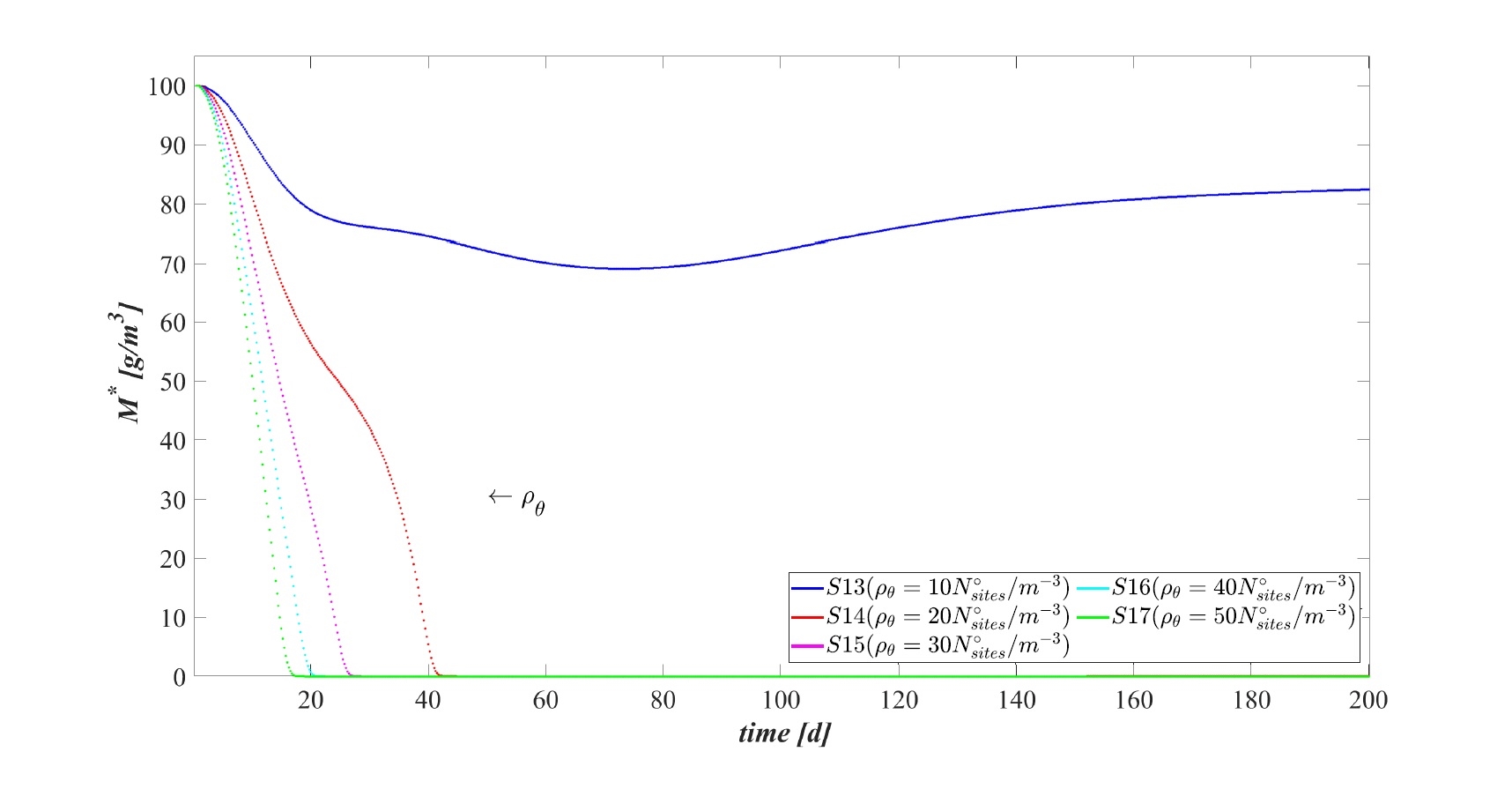}}
\caption{$SET3$ - Evolution of metal concentration within the reactor for different densities of binding sites $\rho_{\theta}$. Wastewater influent composition: $S^{in}_{IC} = 180 \ g \ m^{-3}$ (inorganic carbon), $S^{in}_{DOC} = 500 \ g \ m^{-3}$ (organic carbon), $S^{in}_{NH_3} = 50 \ g \ m^{-3}$ (ammonia), $S^{in}_{NO_3} = 0$ (nitrate), $S^{in}_{O_2} = 0$ (oxygen). $S13: \rho_{\theta}=10 \ N^{\circ}_{sites} \ m^{-3}$, $S14: \rho_{\theta}=20 \ N^{\circ}_{sites} \ m^{-3}$, $S15: \rho_{\theta}=30 \ N^{\circ}_{sites} \ m^{-3}$, $S16: \rho_{\theta}=40 \ N^{\circ}_{sites} \ m^{-3}$, $S17: \rho_{\theta}=50 \ N^{\circ}_{sites} \ m^{-3}$. Incident light intensity: $I_0= 0.008 \ kmol \ m^{-2} \ d^{-1}$. Duration of the cycle: $\tau =6 \ h$. Time of light exposure: $t_{light} = 50$\% $\tau$.} \label{f5.4.3.4}
\end{figure*}

The five values of $\rho_{\theta}$ used are: $10$, $20$, $30$, $40$, $50$ $N^{\circ}_{sites} \ m^{-3}$. The concentration of soluble substrates in the influent wastewater $S^{in}_j$ and initial concentration of planktonic biomasses within the reactor $\psi^*_{i,0}$ set for this numerical study are the same as in $SET1$. Numerical results are summarized in Figs. \ref{f5.4.3.1}-\ref{f5.4.3.4}.  

The time evolution of the granule radius $R(t)$ is shown in Fig. \ref{f5.4.3.1}. It is clear that different densities of binding sites $\rho_{\theta}$ affects the granule evolution in the second stage of the process, since the further radius increment around $20$-$50$ days is associated to the phototrophs growth and phototrophs and $EPS$ have better adsorption capabilities. Indeed, when $\rho_{\theta}$ increases, the granulation process occurs rapidly and the granule reaches the steady-state size quickly. However, such steady-state size is not dependent on the binding sites density. Indeed, the profiles of $R(t)$ get closer over time and reach the same steady-state value, except for $\rho_{\theta}=10 \ N^{\circ}_{sites} \ m^{-3}$. It leads to conclude that with very low densities of binding sites the granulation and adsorption process do not completely evolve.

Fig. \ref{f5.4.3.2} reports the sessile mass of the different microbial species within the granule. Again, relevant differences concern phototrophs growth. The phototrophs mass increases faster when the algal-bacterial consortium have higher binding sites densities. Metal removal and phototrophs growth positively influence each other. A faster metal adsorption enhances the phototrophic growth rate, and phototrophs contribute to accelerate the metal removal process thanks to their high adsorption properties.

This is visible also in Fig. \ref{f5.4.3.3}, where the phototrophic sessile biomass within the granule is shown at different times. Again, relevant differences concern the time frame which goes from $20$ to $50$ days. By increasing $\rho_{\theta}$, phototrophic sessile biomass grows faster and the steady-state microbial distribution is reached earlier. However, the steady-state distribution is the same for all values of densities, except for the simulation $S13$, in which no phototrophic biomass is detected throughout the granule.

Fig. \ref{f5.4.3.6} presents the steady-state configuration of the residual free binding sites within the biofilm granule. Obviously, for $\rho_{\theta}$ equal to $10 \ N^{\circ}_{sites} \ m^{-3}$ the adsorption process is not completed ($S13$), the granule does not completely develop due to the stronger inhibiting effect, and the binding sites are almost occupied. On the contrary, for higher values of binding sites density an increasing residual fraction of adsorption sites can be observed ($S14-S17$). 

Fig. \ref{f5.4.3.5} and \ref{f5.4.3.4} show the trend of soluble substrates and metal concentrations in the SBR effluent, respectively. Each point represents the concentrations of substrates and metal in the effluent at the end of each cycle. As mentioned before, in the initial stage of the process the consumption and production of soluble substrates mainly depend on the metabolic activity of heterotrophic biomass. Consequently, the trends of soluble substrates are not affected by the variation of $\rho_{\theta}$, since the role of heterotrophic bacteria in the adsorption process is marginal. For later times, phototrophic biomass starts to grow, and the trend of substrates (Fig. \ref{f5.4.3.5}) and metal (Fig. \ref{f5.4.3.4}) becomes more sensitive to $\rho_{\theta}$. For high values of $\rho_{\theta}$, the concentrations of soluble substrates and metal achieve the steady-state values earlier and the time required to completely adsorb the residual metal decreases. For $\rho_{\theta}=10 \ N^{\circ}_{sites} \ m^{-3}$, the metal adsorption process does not complete, because of the absence of phototrophs and the small amount of $EPS$ observed throughout the granule.

From the numerical results, it is clear that the density of binding sites $\rho_{\theta}$ influences the adsorption process rate of the free metal, and, therefore, the time necessary for the metal removal. Anyway, it can be concluded that the steady-state configuration in terms of biofilm dimension, microbial species stratification, and metal removal efficiency of the process are not affected by $\rho_{\theta}$ above a critical value. 

\section{Discussion and conclusions} \label{n5.5}
\
Biosorption is proving to be a promising alternative to conventional methods for the removal of metals from municipal and industrial effluents, as microorganisms and their derived products have high biosorption capabilities of inorganic compounds. Indeed, conventional physico/chemical methods for metals removal are expensive and inefficient for very low metals concentrations \cite{abbas2014biosorption, ahluwalia2007microbial, anjana2007biosorption, chojnacka2010biosorption}. Biosorption offers several advantages including cost effectiveness, high efficiency, minimization of chemical compounds utilization, and regeneration of biosorbents \cite{abbas2014biosorption}. Nevertheless, there are practical limitations as living biomass is very sensitive to high metal concentrations \cite{munoz2006sequential}. An understanding of metal toxicity effects in biofilms is crucial to the successfully design bioreactors for the contextual removal of organic contaminants and metals. The mathematical model proposed in this work allows to simulate the formation and evolution of oxygenic photogranules within a granular-based sequencing batch reactor and describing the adsorption process of metals on the matrix of biofilm granules. The most interesting observations resulting from the numerical studies are summarized below:

\begin{itemize}
\item  The adsorption process on oxygenic photogranules matrix shows high removal efficiency. These numerical result is in accordance with experimental works in which more than $99$\% of metal present in aqueous solutions is absorbed using algal-bacterial granules, thanks to their excellent adsorption capacities \cite{yang2015lipid,yang2020enhanced, yang2021insight}.

\item The results outline the key role of phototrophs and $EPS$ in the metal removal process, as phototrophs are good biosorbents and metals stimulate the production of $EPS$ in greater amount and with higher adsorption capabilities. These results reflect what has been observed in Yang et al. \cite{yang2020enhanced}, where a comparison between conventional bacterial granules and algal-bacterial photogranules is performed, demonstrating that algal-bacterial granular biofilms show advantages in both biosorption capacity and granular stability.
 
\item Furthermore, the model confirms that the performances of the adsorption process can be significantly affected by the metals concentration present in the wastewater. The highest removal efficiencies are achieved for low concentrations of metal in the influent \cite{abbas2014biosorption}. Indeed, higher is the metal concentration in the influent and stronger will be the inhibiting effect on the microbial growth. Although $EPS$ content significantly increases in presence of metals \cite{yang2015lipid}, numerical results show that it is not sufficient in case of very high metal concentration. Moreover, as shown by Yang et al. \cite{yang2015lipid}, biomass growth is not or is little inhibited by certain concentrations of heavy metals, confirming that algal biomass could efficiently remove them through intracellular accumulation and extracellular immobilization.

\item Lastly, the model results show how a higher density of binding sites, induced by heat or acids pretreatments, may enhance the adsorption process and reduce the time required for the complete degradation of substrates and removal of metals \cite{ahalya2003biosorption}.
\end{itemize}

Most of the results shown are qualitatively in accordance with the experimental evidence reported in literature. Accordingly, this model is able to correctly simulate both the formation and maturation of oxygenic photogranules and removal process of toxic metals. From an engineering point of view, this allows to conclude that the model represents a useful tool in studying the removal processes of both organic and inorganic compounds in granular-based sequencing batch reactor systems. 

 The present work demonstrates the potential applicability of the algal-bacterial granules towards removal of more than one heavy metal. Nevertheless, their joint removal could be not as easy as the removal of a single contaminant. This could be caused by antagonistic effects between the different metals. Looking forward, research activities should be geared towards ways to minimize the antagonistic effects between contaminants.

\clearpage
\bibliographystyle{unsrt}      
\bibliography{Bibliography}

\end{document}